\documentclass[preprint]{raa}   
\usepackage{graphicx,times}             %
\usepackage{natbib}
\usepackage{amssymb,amsmath}
\usepackage{longtable}
\bibpunct{(}{)}{;}{a}{}{,}

\newcommand{\wyAdd}{}
\newcommand{\orcid}[1]{%
    \raisebox{0.7ex}{\scalebox{1}{%
        \href{https://orcid.org/#1}{\includegraphics[height=1.5ex]{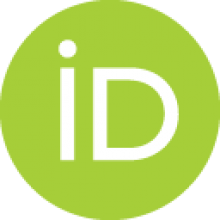}}%
    }}%
}

\newcommand{\feii}{\mbox{Fe\,{\sc II}}}
\newcommand{\mgi}{\mbox{Mg\,{\sc I}}}
\newcommand{\mgii}{\mbox{Mg\,{\sc II}}}

\newcommand{\nai}{\mbox{Na\,{\sc I}}}
\newcommand{\caii}{\mbox{Ca\,{\sc II}}}
\newcommand{\cai}{\mbox{Ca\,{\sc I}}}
\newcommand{\mnii}{\mbox{Mn\,{\sc II}}}
\newcommand{\oii}{\mbox{O\,{\sc II}}}

\usepackage[pagebackref=true]{hyperref}

\usepackage{placeins} 

\begin{document}

\title{GRB 240825A: Early Reverse Shock and Its Physical Implications}

   \volnopage{Vol.0 (20xx) No.0, 000--000}      %
   \setcounter{page}{1}          %
   
\author{
Chao Wu\orcid{0009-0001-7024-3863}$^*$
\inst{1,2}
\and Yun Wang\orcid{0000-0002-8385-7848}$^*$
\inst{3}
\and Hua-Li Li
\inst{1}
\and Li-Ping Xin\orcid{0000-0002-9422-3437}
\inst{1,2}
\and Dong Xu\orcid{0000-0003-3257-9435}
\inst{1}
\and Benjamin Schneider\orcid{0000-0003-4876-7756 }
\inst{4,5}
\and Antonio de Ugarte Postigo\orcid{0000-0001-7717-5085}
\inst{6}
\and Gavin Lamb\orcid{0000-0001-5169-4143}
\inst{7}
\and Andrea Reguitti\orcid{0000-0003-4254-2724}
\inst{8,9}
\and Andrea  Saccardi\orcid{0000-0002-6950-4587}
\inst{10}
\and Xing Gao
\inst{11}
\and Xing-Ling Li
\inst{12}
\and Qiu-Li Wang\orcid{0009-0000-9352-6447}
\inst{3,13}
\and Bing Zhang\orcid{0000-0002-9725-2524}
\inst{14,15}
\and Jian-Yan Wei
\inst{1,2}
\and Shuang-Nan Zhang
\inst{16,17}
\and Fr\'ed\'eric Daigne
\inst{18}
\and  Jean-Luc Atteia\orcid{0000-0001-7346-5114}
\inst{19}
\and Maria-Grazia Bernardini\orcid{0000-0001-6106-3046}
\inst{8}
\and Hong-bo Cai
\inst{1}
\and Arnaud  Claret
\inst{20}
\and Bertrand Cordier
\inst{20}
\and Jin-Song  Deng 
\inst{1,2}
\and Olivier Godet
\inst{19}
\and Diego G\"{o}tz\orcid{0000-0001-9494-0981}
\inst{10}
\and Xu-Hui Han\orcid{0000-0002-6107-0147}
\inst{1}
\and Zhe Kang
\inst{21}
\and Guang-Wei Li
\inst{1,2}
\and Zhen-Wei Li
\inst{21}
\and Cheng-Zhi Liu
\inst{21}
\and Xiao-Meng Lu
\inst{1}
\and You Lv
\inst{21}
\and Julian P  Osborne\orcid{0000-0002-1041-7542}
\inst{22}
\and Jesse T. Palmerio\orcid{0000-0002-9408-1563}
\inst{20}
\and Yu-Lei Qiu
\inst{1}
\and St\'{e}phane Schanne\orcid{0009-0007-1271-9900}
\inst{20}
\and Damien Turpin\orcid{0000-0003-1835-1522}
\inst{10}
\and Susanna Diana Vergani\orcid{0000-0001-9398-490}
\inst{23}
\and Jing Wang
\inst{1,2}
\and Yu-Jie Xiao
\inst{1}
\and Wen-Jin Xie
\inst{1}
\and Yang Xu
\inst{1}
\and Zhu-Heng Yao\orcid{0009-0000-1228-2373}
\inst{1}
\and Pin-Pin Zhang
\inst{1}
\and Ruo-Son Zhang
\inst{1}
\and Cheng-Wei Zhu
\inst{21}
\and  Riccardo Brivio\orcid{0009-0000-0564-7733}
\inst{8}
\and Stefano Covino\orcid{0000-0001-9078-5507}
\inst{8,24}
\and Paolo D'Avanzo\orcid{0000-0001-7164-1508}
\inst{8}
\and Matteo Ferro\orcid{0009-0007-5708-7978}
\inst{8}
\and Andrea Melandri\orcid{0000-0002-2810-2143}
\inst{25}
\and Andrea Rossi\orcid{0000-0002-8860-6538}
\inst{26}
\and  Jos\'e Feliciano Ag\"u\'i Fern\'andez\orcid{0000-0001-6991-7616}
\inst{27}
\and Christina C. Th\"one\orcid{0000-0002-7978-7648}
\inst{28}
\and  Chun-Hai Bai
\inst{11}
\and Ali Esamdin
\inst{11}
\and Abdusamatjan Iskandar
\inst{11}
\and Shahidin Yaqup
\inst{11}
\and Yu Zhang
\inst{11}
\and Tu-Hong Zhong
\inst{11}
\and  Shao-Yu Fu\orcid{0009-0002-7730-3985}
\inst{1}
\and Shuai-Qing Jiang\orcid{0009-0001-8155-7905}
\inst{1,2}
\and Xing Liu
\inst{1,2}
\and  Jie An\orcid{0009-0000-5068-3434}
\inst{1,2}
\and Zi-Pei Zhu\orcid{0000-0002-9022-1928}
\inst{1}
\and  Jia-Xin Cao
\inst{12}
\and En-Wei Liang\orcid{0000-0002-7044-733X}
\inst{12}
\and Da-Bin Lin\orcid{0000-0003-1474-293X}
\inst{12}
\and Xiang-Gao Wang\orcid{0000-0001-8411-8011}
\inst{12}
\and  Guo-Wang Du
\inst{29}
\and Xin-Zhong Er
\inst{29}
\and Yuan Fang
\inst{29}
\and Xiao-Wei Liu
\inst{29}
\and Christophe Adami
\inst{30}
\and Michel Dennefeld 
\inst{31}
\and Emeric  Le Floc'h
\inst{10}
\and  Johan Peter Uldall  Fynbo\orcid{0000-0002-8149-8298}
\inst{32,33}
\and P\'all  Jakobsson
\inst{34}
\and Daniele Bj$\o$rn Malesani\orcid{0000-0002-7517-326X}
\inst{32,33,35}
\and  Zhi-Ping Jin\orcid{0000-0003-4977-9724}
\inst{3,13}
\and Jia Ren\orcid{0000-0002-9037-8642}
\inst{3}
\and Hao Wang\orcid{0000-0002-0556-1857}
\inst{3}
\and Da-Ming Wei\orcid{0000-0002-9758-5476}
\inst{3,13}
\and Hao Zhou\orcid{0000-0003-2915-7434}
\inst{3}
\and  Sergio Campana\orcid{0000-0001-6278-1576}
\inst{8}
\and Shiho  Kobayashi \orcid{0000-0001-7946-4200}
\inst{36}
\and Massimiliano De  Pasquale\orcid{0000-0002-4036-7419}
\inst{37}
}
\institute{
National Astronomical Observatories, Chinese Academy of Sciences, Beijing 100101,  China; \it{xlp@nao.cas.cn,wjy@nao.cas.cn} \\ 
\and 
School of Astronomy and Space Science, University of Chinese Academy of Sciences, Beijing 101408, China \\ 
\and 
Purple Mountain Observatory, Chinese Academy of Sciences, Nanjing 210023, China; \it{wangyun@pmo.ac.cn} \\ 
\and 
Aix Marseille University, CNRS, CNES, LAM, Marseille, France \\ 
\and 
Massachusetts Institute of Technology, Kavli Institute for Astrophysics and Space Research, Cambridge, Massachusetts, USA \\ 
\and 
Aix Marseille Univ, CNRS, CNES, LAM Marseille, France \\ 
\and 
Astrophysics Research Institute, Liverpool John Moores University, IC2 Liverpool Science Park, 146 Brownlow Hill, Liverpool, L3 5RF, UK \\ 
\and 
INAF - Osservatorio Astronomico di Brera, via E. Bianchi 46, I-23807 Merate (LC), Italy \\ 
\and 
INAF – Osservatorio Astronomico di Padova, Vicolo dell’Osservatorio 5, I-35122 Padova, Italy \\ 
\and 
Universit\'e Paris-Saclay, Universit\'e Paris Cit\'e, CEA, CNRS, AIM, 91191, Gif-sur-Yvette, France \\ 
\and 
Xinjiang Astronomical Observatory, Chinese Academy of Sciences, Urumqi, Xinjiang, 830011, China \\ 
\and 
Guangxi Key Laboratory for Relativistic Astrophysics, School of Physical Science and Technology, Guangxi University, 530004 Nanning, Guangxi, China \\ 
\and 
School of Astronomy and Space Science, University of Science and Technology of China, Hefei 230026, China \\ 
\and 
Nevada Center for Astrophysics, University of Nevada, Las Vegas, NV 89154, USA \\ 
\and 
Department of Physics and Astronomy, University of Nevada, Las Vegas, NV 89154, USA \\ 
\and 
Key Laboratory of Particle Astrophysics, Institute of High Energy Physics, Chinese Academy of Sciences, Beijing 100049, China \\ 
\and 
University of Chinese Academy of Sciences, Chinese Academy of Sciences, Beijing 100049, China \\ 
\and 
Sorbonne Universit\'e, CNRS, UMR 7095, Institut d'Astrophysique de Paris, 98 bis bd Arago, F-75014 Paris, France \\ 
\and 
IRAP, Universit\'e de Toulouse, CNRS, CNES, Toulouse, France \\ 
\and 
CEA Paris-Saclay, Irfu/Département d'Astrophysique, 9111 Gif sur Yvette, France \\ 
\and 
Changchun Observatory, National Astronomical Observatories, Chinese Academy of Sciences,Changchun 130117, China \\ 
\and 
School of Physics and Astronomy, University of Leicester, University Road, Leicester, LE1 7RH, UK \\ 
\and 
LUX, Observatoire de Paris, PSL University, CNRS, Sorbonne University, 92190 Meudon, France \\ 
\and 
Como Lake centre for AstroPhysics (CLAP), DiSAT, Università dell’Insubria, via Valleggio 11, 22100 Como, Italy \\ 
\and 
INAF - Osservatorio Astronomico di Roma, Via di Frascati 33, I-00040, Monte Porzio Catone (RM), Italy \\ 
\and 
INAF – Osservatorio di Astrofisica e Scienza dello Spazio, via Piero Gobetti 93/3, I-40024, Bologna, Italy \\ 
\and 
Centro Astron\'omico Hispano en Andaluc\'ia, Observatorio de Calar Alto, Sierra de los Filabres, G\'ergal, Almer\'ia, 04550, Spain \\ 
\and 
E. Kharadze Georgian National Astrophysical Observatory, Mt. Kanobili, Abastumani 0301, Adigeni, Georgia \\ 
\and 
South-Western Institute for Astronomy Research, Yunnan University, Kunming, Yunnan 650504, People's Republic of China \\ 
\and 
Marseille Univ., CNRS, CNES, LAM, 13388 Marseille, France \\ 
\and 
Sorbonne Universit\'e, CNRS, UMR7095, Institut d'Astrophysique de Paris, 98bis Bd Arago, 75014 Paris,France \\ 
\and 
Niels Bohr Institute, University of Copenhagen, Jagtvej 155, 2200, Copenhagen N, Denmark \\ 
\and 
The Cosmic Dawn Centre (DAWN), Denmark \\ 
\and 
Centre for Astrophysics and Cosmology, Science Institute, University of Iceland, Dunhagi 5, 107 Reykjavík, Iceland \\ 
\and 
Department of Astrophysics/IMAPP, Radboud University, PO Box 9010, 6500 GL, The Netherlands \\ 
\and 
Astrophysics Research Institute, Liverpool John Moores University, 146 Brownlow Hill, Liverpool L3 5RF, United Kingdom \\ 
\and 
Department of Mathematics and Computer Sciences, Physical Sciences and Earth Sciences of University of Messina, Papardo Campus, Via F. S. D'Alcontres 31, 98166 Messina Italy \\ 
\vs\no
   {\small Received 20xx month day; accepted 20xx month day}}

\abstract{ 
Early multiwavelength observations offer crucial insights into the nature of the relativistic jets responsible for gamma-ray bursts and their interaction with the surrounding medium.
We present data of GRB 240825A from 17 space- and ground-based telescopes/instruments, covering wavelengths from NIR/optical to X-ray and GeV, and spanning from the prompt emission to the afterglow phase triggered by Swift and Fermi.
The early afterglow observations were carried out by SVOM/C-GFT, and spectroscopic observations of the afterglow by GTC, VLT, and TNG determined the redshift of the burst ($z = 0.659$) later.
A comprehensive analysis of the prompt emission spectrum observed by Swift-BAT and Fermi-GBM/LAT reveals a rare and significant high-energy cutoff at ~76 MeV. Assuming this cutoff is due to  $\gamma\gamma$ absorption allows us to place an upper limit on the initial Lorentz factor, $\Gamma_0 < 245$. The optical/NIR and GeV afterglow light curves can be described by the standard external shock model, with early-time emission dominated by a reverse shock (RS) and a subsequent transition to forward shock (FS) emission. Our afterglow modeling yields a consistent estimate of the initial Lorentz factor ($\Gamma_{\rm 0} \sim 234$). Furthermore, the RS-to-FS magnetic field ratio ($\mathcal{R}_B \sim 302$) indicates that the reverse shock region is significantly more magnetized than the FS region. An isotropic-equivalent kinetic energy of $E_{\text{k,iso}} = 5.25 \times 10^{54}$ erg is derived, and the corresponding $\gamma$-ray radiation efficiency is estimated to be $\eta_{\gamma}$ = 3.1\%. On the other hand, the standard afterglow model cannot reproduce the X-ray light curve of GRB 240825A, calling for improved models to characterize all multiwavelength data.
\keywords{ (stars:) gamma-ray burst: individual:(GRB 240825A)\textendash (stars:) gamma-ray burst: general\textendash (transients:) gamma-ray bursts}
}
\def\thefootnote{*}\footnotetext{These authors contributed equally to this work}\def\thefootnote{\arabic{footnote}}
\def\thefootnote{}\footnotetext{Corresponding authors:Yun Wang (wangyun@pmo.ac.cn), Li-Ping Xin (xlp@nao.cas.cn), Jian-Yan Wei (wjy@nao.cas.cn) }\def\thefootnote{\arabic{footnote}}
\authorrunning{C. Wu, Y. Wang, \& H. -L. Li et al. }            %
\titlerunning{GRB 240825A: Reverse Shock \& Physics }  %
\maketitle

\section{Introduction} 
\label{sec:intro}
Gamma-ray bursts (GRBs) are the most luminous explosions observed in the universe. Over more than half a century of research in this field, the cosmological origins and the demands of extreme relativistic conditions have been acknowledged as fundamentally significant \citep{1975NYASA.262..164R,1997Natur.387..878M}. The GRB jet, which is either Poynting-flux-dominated or matter-dominated, undergoes acceleration through the dissipation of its magnetic or thermal energy and reaches relativistic velocities. For prompt emission, the physical origin of the initial emission, i.e. the prompt phase, remains debated, with several scenarios proposed to explain it, such as internal shocks \citep[e.g.][]{1994ApJ...430L..93R,1994ApJ...427..708P,1997ApJ...490...92K,1998MNRAS.296..275D}, magnetic dissipation \citep[e.g.][]{2001A&A...369..694S,2002A&A...391.1141D,2008A&A...480..305G,2011ApJ...726...90Z,2012MNRAS.419..573M,2015MNRAS.453.1820K,2016MNRAS.462...48S,2016MNRAS.459.3635B,2016ApJ...816L..20G}, and photospheric emission \citep[e.g.][]{1986ApJ...308L..47G,1994MNRAS.270..480T,1999ApJ...511L..93G,2006ApJ...642..995P,2007ApJ...666.1012T}, etc. The afterglow originates from external dissipation, specifically external shocks sweeping up mass from the circumburst medium, which may include both forward shock (FS) and reverse shock (RS) components \citep{1997ApJ...476..232M,1999MNRAS.306L..39M,1999ApJ...517L.109S}. Therefore, complete observations from prompt emission to afterglow are particularly crucial for understanding the physics of GRBs, including the evolution of their Lorentz factors, the properties of their surrounding environments, and other related aspects. \wyAdd{Among them, the relatively rare optical components arising from the RS can be used as probes to investigate the physics of GRB jets} \citep{2002ChJAA...2..449F,2003ApJ...595..950Z}, particularly in constraining the magnetization of the FS and RS regions \citep{2004A&A...424..477F,2005ApJ...628..315Z}.

In this paper, we present GRB 240825A, a GRB with extensive multiwavelength observations covering both the prompt emission and the afterglow phases. This event was triggered and localized by Swift \citep{2024GCN.37274....1G}. Follow-up observations of the early and late phases of the afterglow were conducted using Chinese Ground Follow-up Telescope (C-GFT) and the onboard instrument Visible Telescope (VT) of the SVOM mission\footnote{\url{https://www.svom.eu/}}. The SVOM mission, dedicated to the study of GRBs, was successfully launched on 2024 June 22 from the Xichang Satellite Launch Center. 
Notably, SVOM/ECLAIRs did not trigger on this burst; the instrument was not collecting data at the time, as it was inside the South Atlantic Anomaly (SAA) of the Earth's magnetic field. Furthermore, the burst was located outside ECLAIRs' field of view (FOV) and, being in Earth occultation, was also not visible to SVOM/GRM.
Additionally, observations from a global network of ground-based telescopes, such as HMT, T80, PAT17, ALT100C, LCOGT, NOT, GTC, VLT, Asiago, REM, TNG, LBT, and MISTRAL@OHP, have contributed to a comprehensive optical photometric and spectral dataset. This multi-facility campaign provides exceptional time-domain and multiband coverage for in-depth afterglow analysis.
In the high-energy prompt emission spectrum of this GRB, a relatively rare high-energy spectral cutoff was detected. According to previous statistical studies \citep{Tang_2015,2023ApJ...943..145L}, only 14 such cases of high-energy spectral cutoffs have been reported.  The high-energy spectral cutoff provides approximate constraints for estimating the initial Lorentz factor \citep{1991ApJ...373..277K,1993A&AS...97...59F,1995ApJ...453..583W,1997ApJ...491..663B,2001ApJ...555..540L,2006ApJ...650.1004B,2011ApJ...729..114A}, offering a valuable comparison to the value derived from  afterglow modeling. Moreover, the well-sampled multiwavelength observations, covering the entire evolution of the afterglow, offer an excellent foundation for deriving the microphysical parameters of this burst. 

Section \ref{sec:observation}  presents a comprehensive overview of the processing and analysis of multiwavelength observational data, spanning $\gamma$-ray, X-ray, and optical bands. Section \ref{sec:af} details the multiwavelength data fitting and the inference of microphysical parameters based on the afterglow model, incorporating updated priors from the prompt emission. Using the inferred results, we compare the parameter  which describes the magnetic ratio between the FS and RS regions, as well as the initial Lorentz factor  with previous statistical studies. Section \ref{sec:sd} provides a summary and discussion of the analysis results for this GRB, covering both the prompt emission and the afterglow phases. Throughout this paper, we adopt the standard cosmological model as $H_{0}=69.6$ km s$^{-1}$Mpc$^{-1}$, $\Omega_{M}=0.29$ and $\Omega_{\Lambda}=0.71$ \citep{2014ApJ...794..135B}. And the errors correspond to a 1 $\sigma$ confidence level by default.

\section{Observations} \label{sec:observation}
GRB 240825A triggered both the Swift Burst Alert Telescope \citep[BAT, ][]{2005SSRv..120..143B} and the Fermi Gamma-Ray Burst Monitor \citep[GBM, ][]{meegan2009fermi} nearly simultaneously at 15:52:59 UT on 2024 25 August, with BAT being triggered 1 second earlier \citep{2024GCN.37274....1G,2024GCN.37273....1F}. Approximately 83 seconds after the trigger, Swift X-Ray Telescope \citep[XRT, ][]{2005SSRv..120..165B} and Ultra-Violet/Optical Telescope \citep[UVOT, ][]{2005SSRv..120...95R} began follow-up observations. Within the XRT error circle of 12.3 arcseconds, a bright optical afterglow was detected at the coordinates RA = 344.57192 deg, DEC = 1.02686 deg\citep{2024GCN.37274....1G}. Notably, SVOM/C-GFT commenced follow-up observations only 66 s after the trigger \citep{2024GCN.37292....1S}, achieving a response time comparable to that of space-based facilities. As such, this event serves as an exemplary case of successful ground-space coordinated observations and embodies a key vision of the SVOM mission \citep{2016arXiv161006892W}.
 The following sections provide a detailed description of the data processing and analysis for the $\gamma$-ray, X-ray, and optical observations.

\subsection{High-energy observations}
\subsubsection{Prompt emission phase}
During the prompt emission phase, the $\gamma$-ray monitors onboard Swift and Fermi (i.e., BAT and GBM) observed GRB 240825A, and the Large Area Telescope \citep[LAT,][]{2009ApJ...697.1071A} on Fermi further detected high-energy photons from this source. We conducted a joint analysis of the detected data from these three instruments. For Swift, BAT is a hard X-ray telescope that uses a coded mask to provide a wide field of view and localization, \wyAdd{and energy band covers 15 keV to 350 keV \citep{2005SSRv..120..143B}.} The data reduction for BAT followed the standard official analysis threads \footnote{\url{https://www.swift.ac.uk/analysis/bat/index.php}} to generate the light curve, spectrum and response files. For Fermi, GBM comprises 12 thallium-activated sodium iodide (NaI(Tl)) scintillation detectors and two bismuth germanate (BGO) scintillation detectors \citep{meegan2009fermi}. Based on the pointing direction of each detector and the position of the source, we selected the data from one NaI detector (n7) and one BGO detector (b1) for the analysis of this event. The Fermi-LAT instrument is a high-performance $\gamma$-ray telescope for the photon energy range from 30 MeV to more than 300 GeV \citep{2009ApJ...697.1071A}. The data reduction for Fermi GBM and LAT was performed using the \texttt{GBM Data Tools v1.1.1} \citep{GbmDataTools} and {\tt Fermitools v2.2.0} package\footnote{\url{https://github.com/fermi-lat/Fermitools-conda/}}, respectively. The GBM response files were obtained from the Fermi FTP server \footnote{\url{https://heasarc.gsfc.nasa.gov/FTP/fermi/}} and are specific to each event, whereas the LAT data were analyzed using the {\tt P8R3\_TRANSIENT020\_V3} instrument response function (IRF).

\begin{figure}[!htp]
    \centering
    \includegraphics[width=0.5\textwidth]{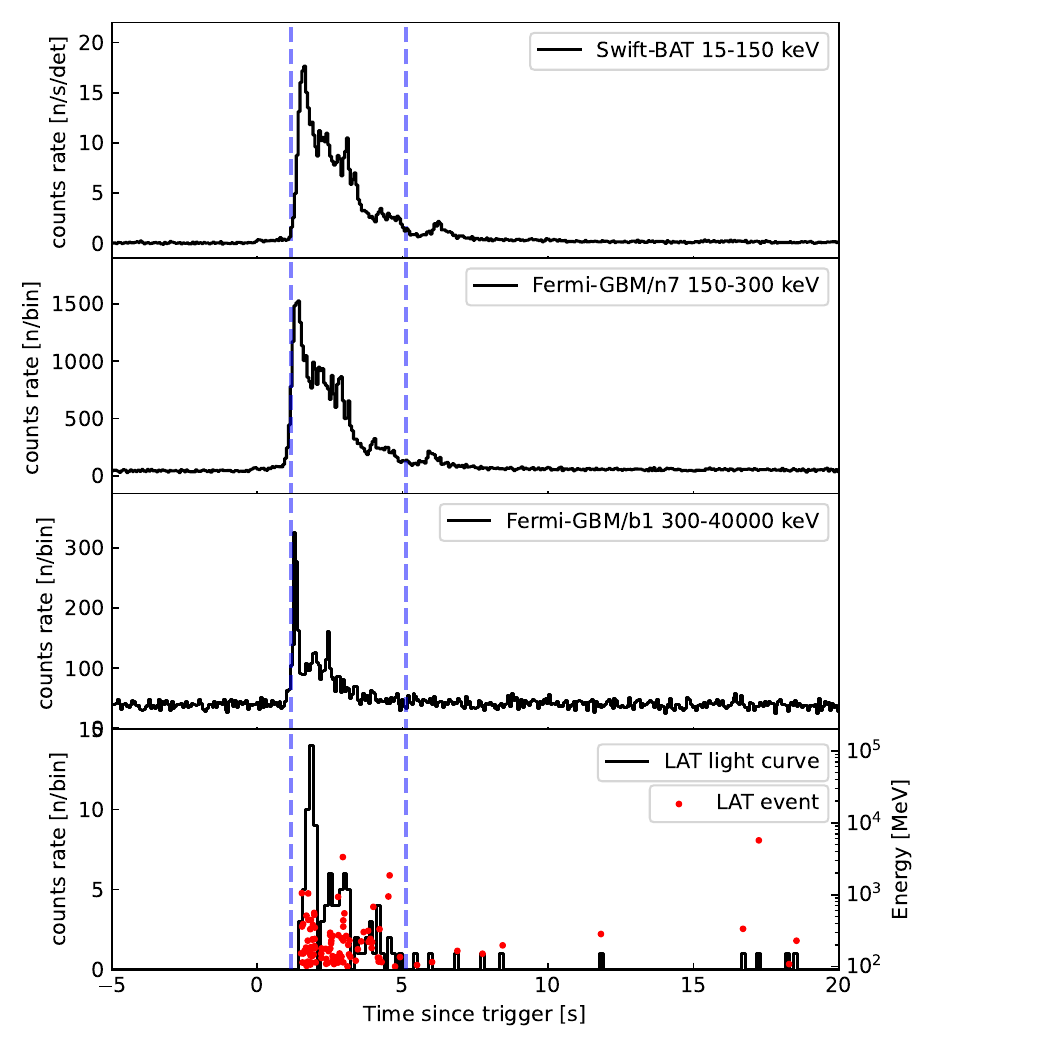}
    \caption{The prompt emission light curve of GRB 240825A was detected by instruments including Swift-BAT and Fermi-GBM/LAT. The blue dashed lines indicate the $T_{90}$ time interval of GBM (50–300 keV), corresponding to $T_0$ + [1.15, 5.12] s \citep{2024GCN.37273....1F}. The black solid lines represent the rebinned light curve, and for the LAT data in the fourth panel, the red points signify the energy of each event.}
    \label{fig:prompt_LC}
\end{figure}

\begin{figure*}[!htp]
    \centering
    \includegraphics[width=0.45\textwidth]{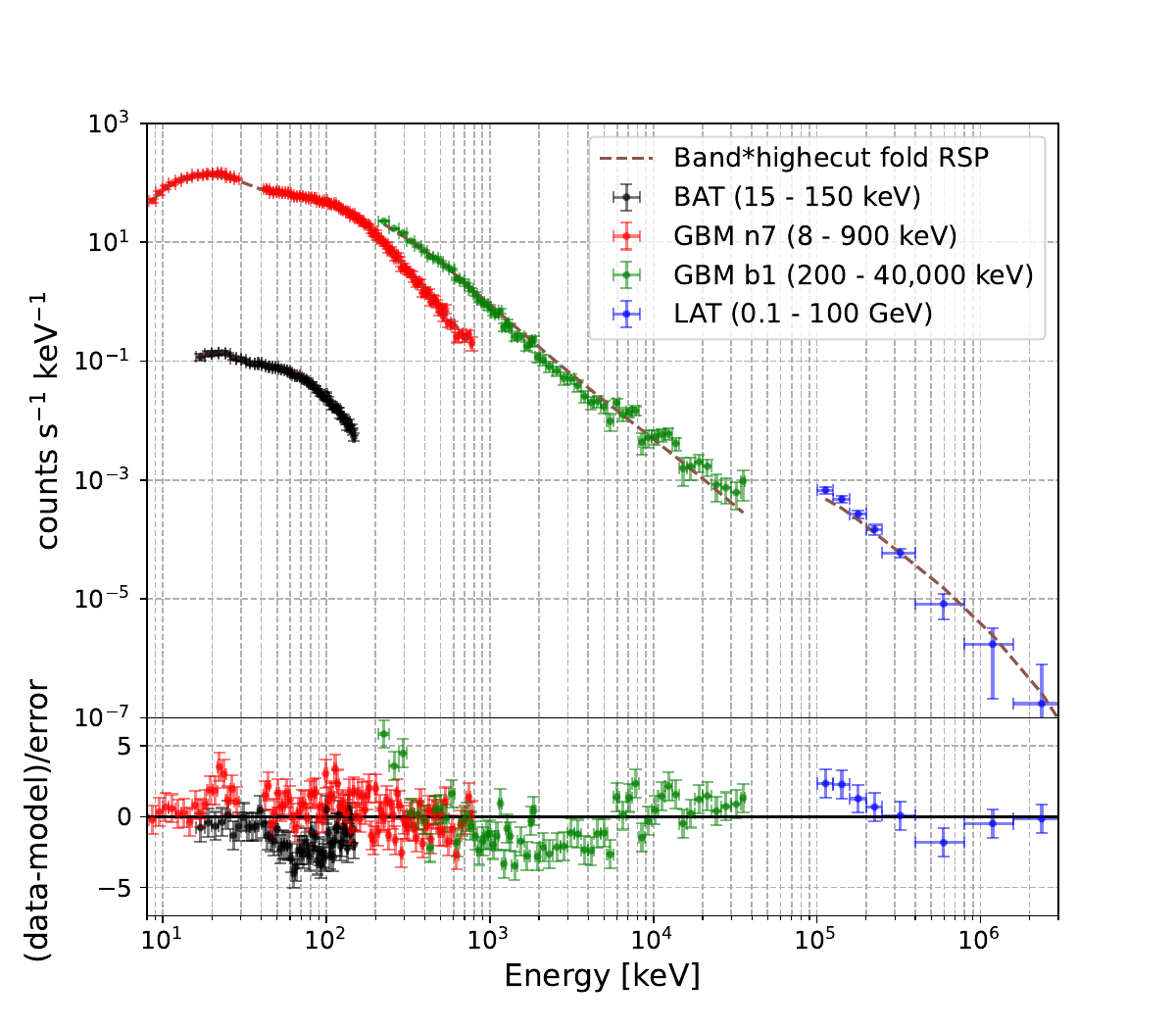}
    \includegraphics[width=0.45\textwidth]{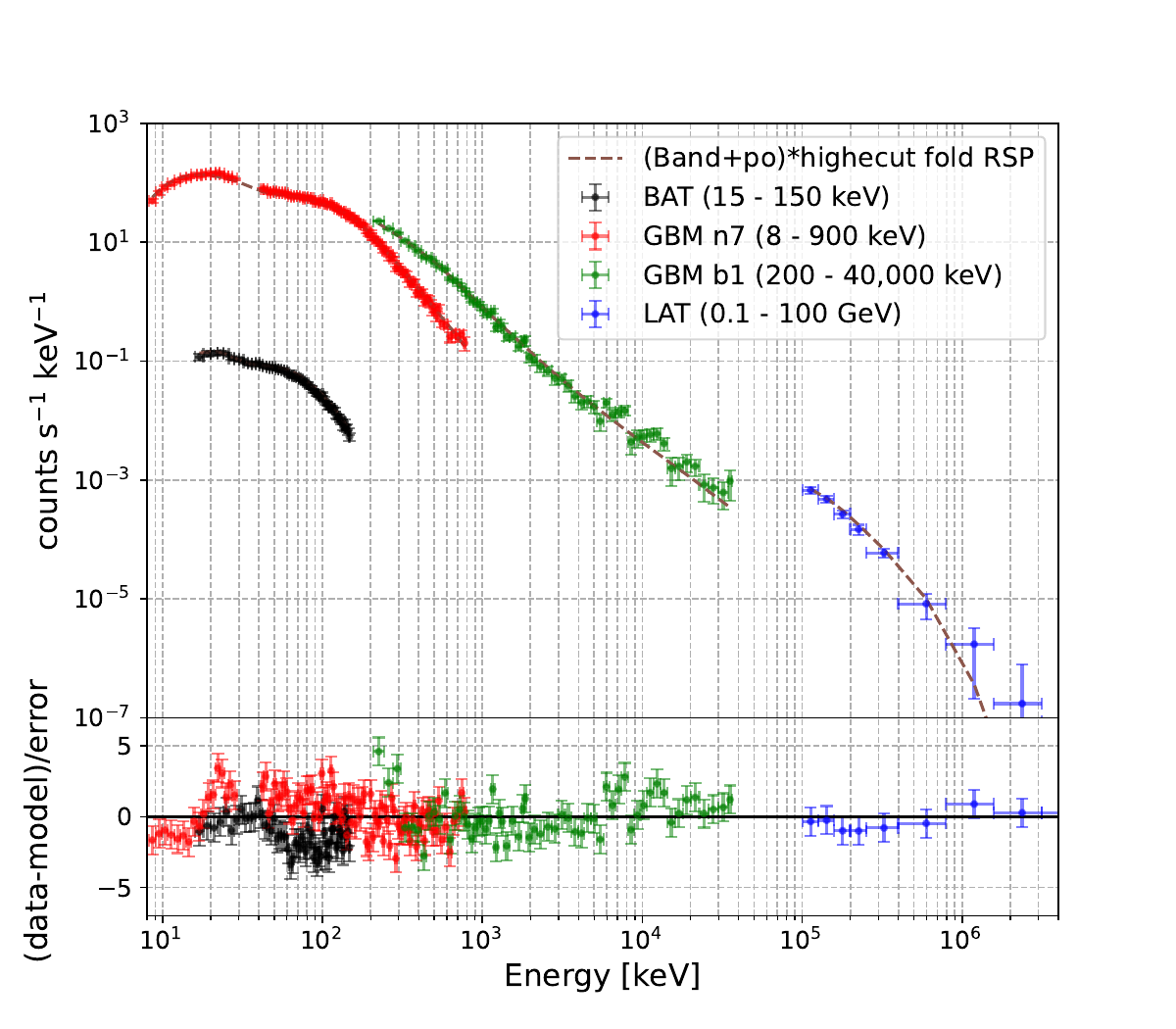}
    \includegraphics[width=0.45\textwidth]{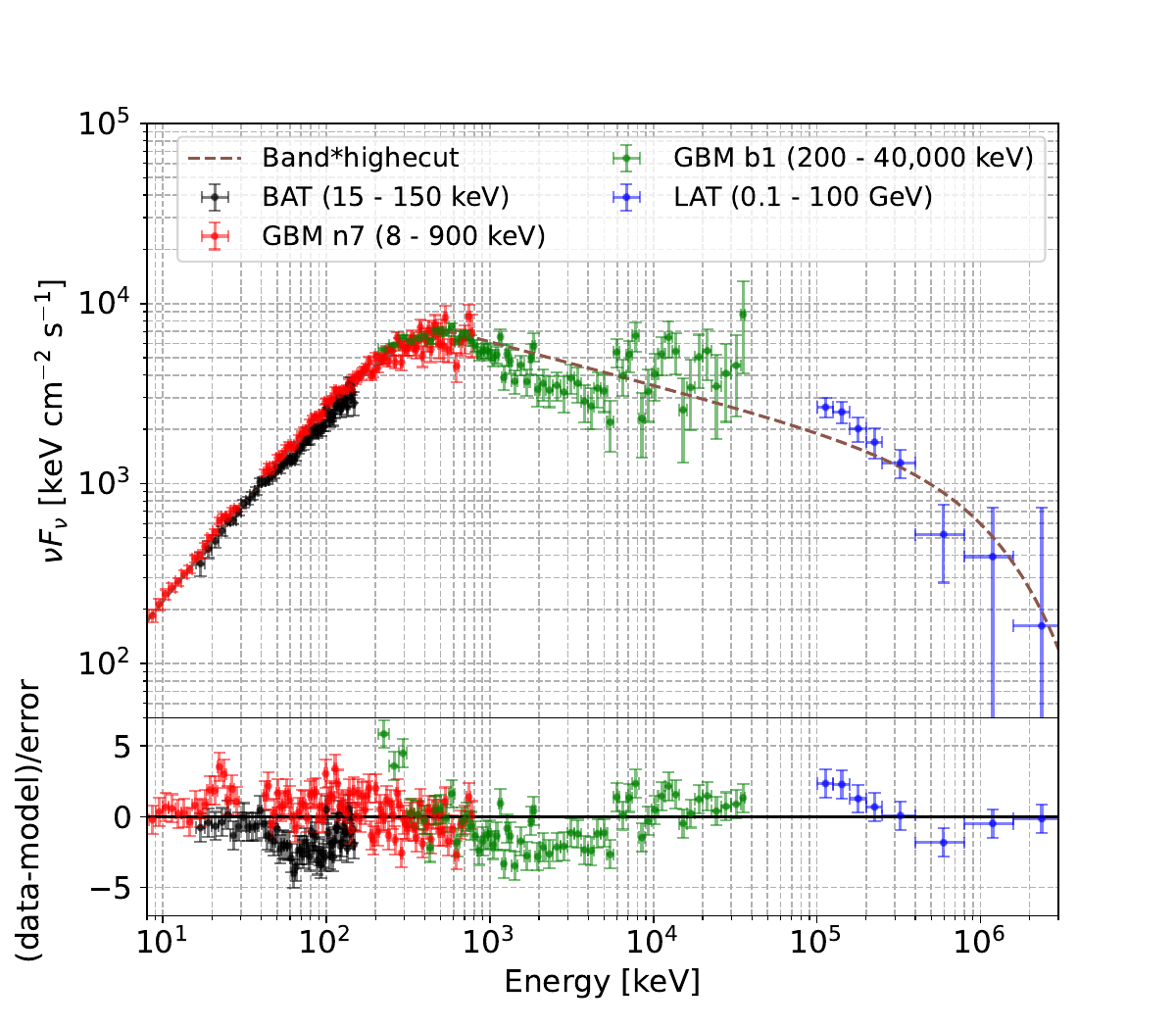}
    \includegraphics[width=0.45\textwidth]{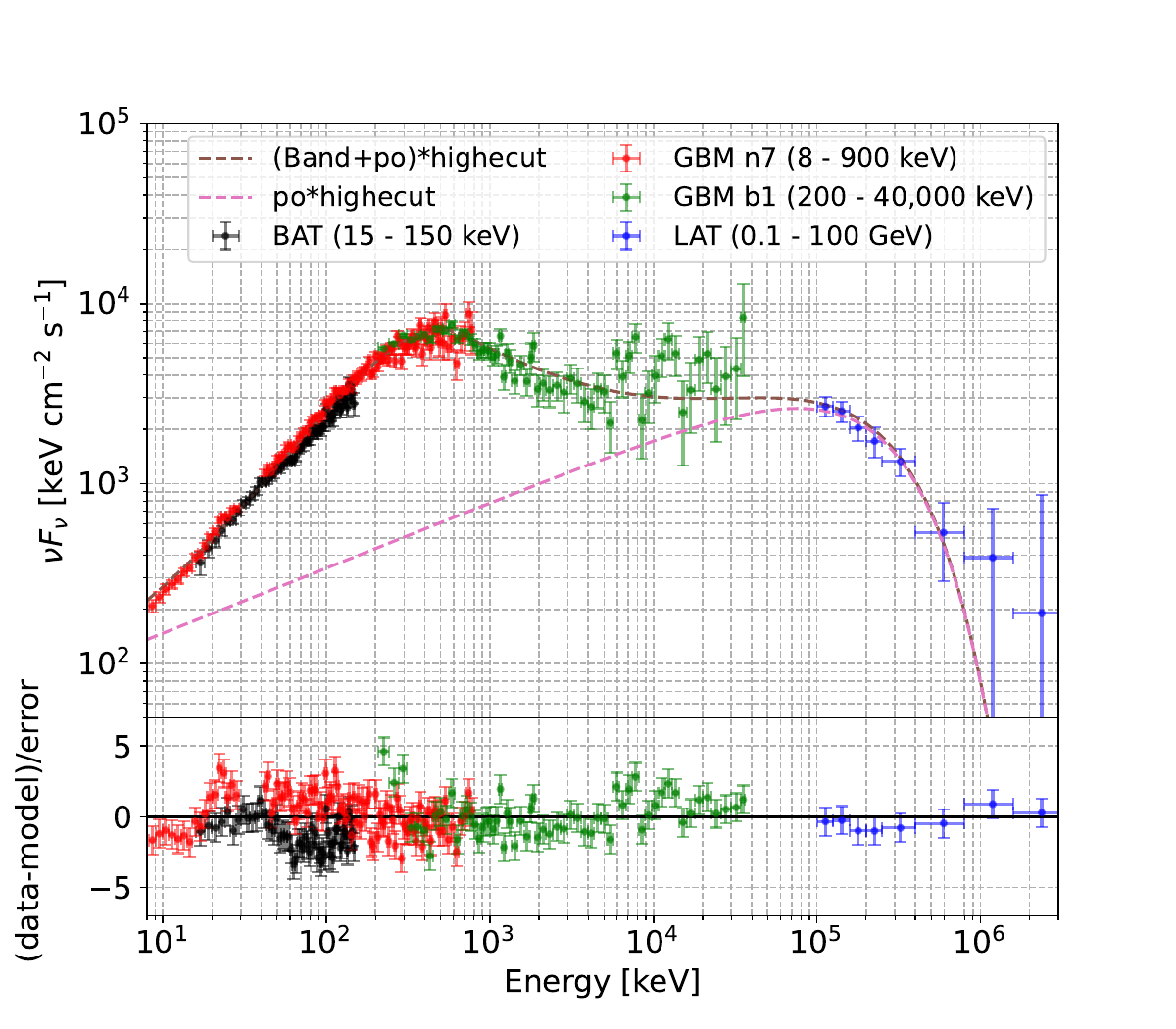}
    \includegraphics[width=0.4\textwidth]{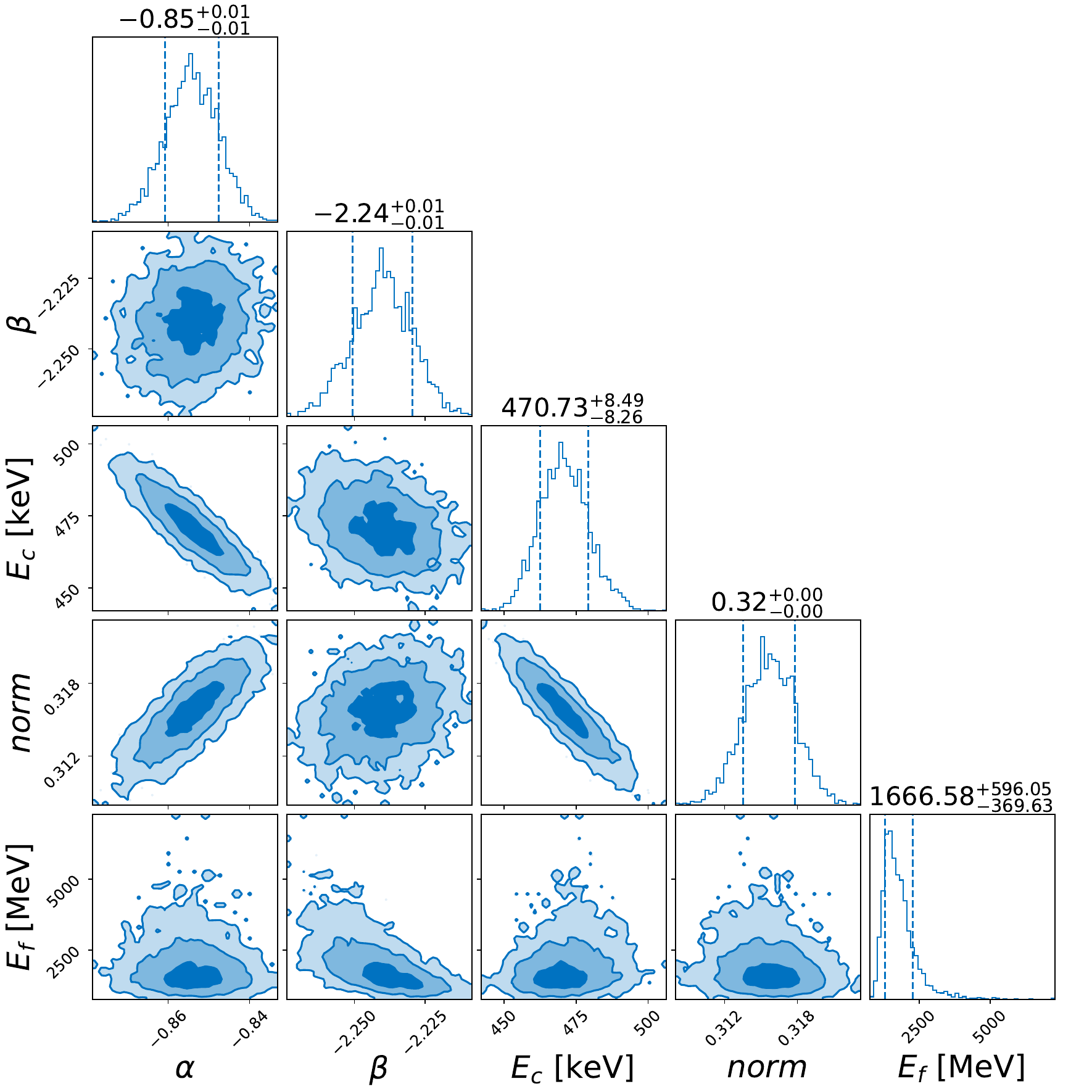}
    \includegraphics[width=0.45\textwidth]{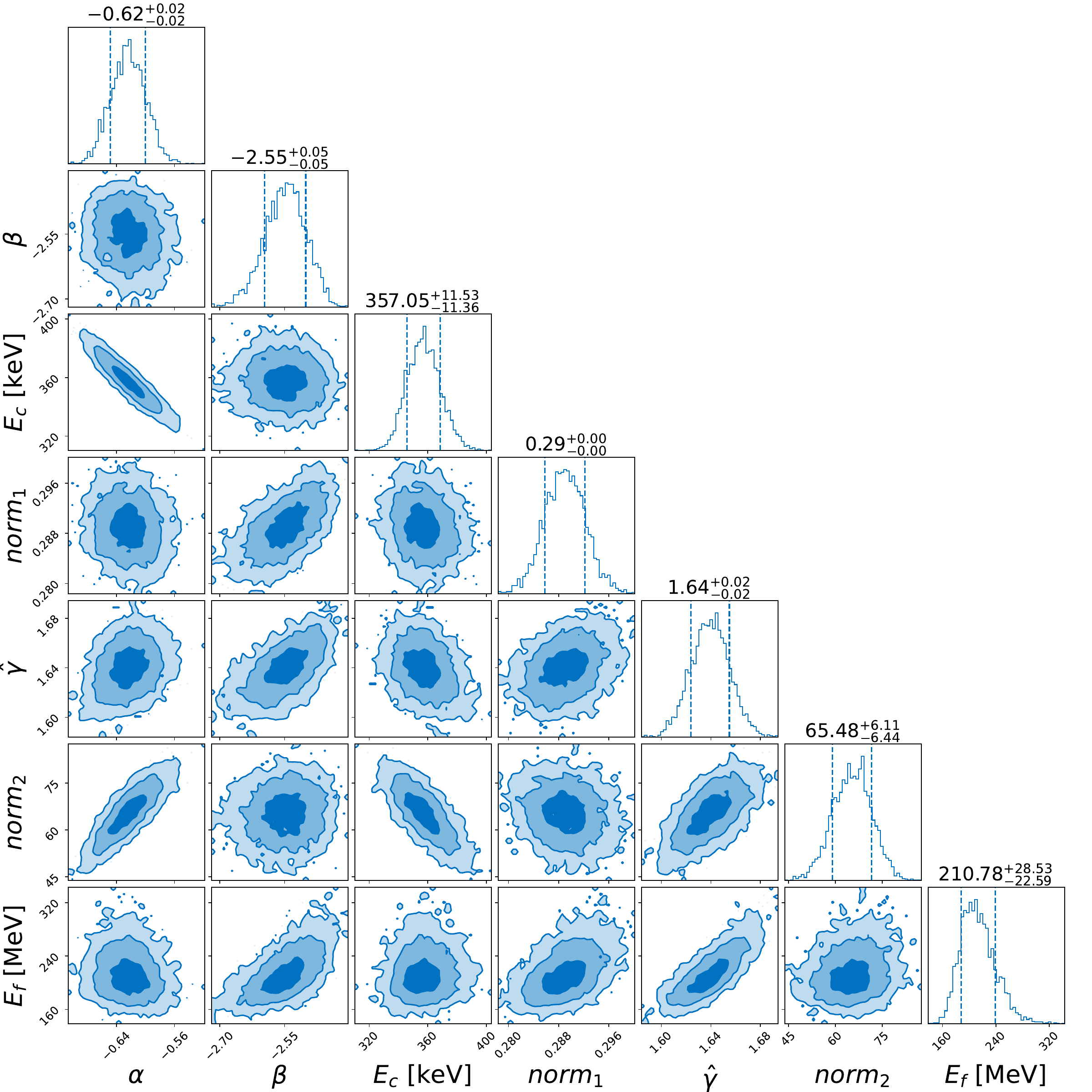}
    \caption{\wyAdd{The fitting results of the time-integrated spectrum of the prompt emission. From top to bottom: the photon spectrum \wyAdd{convoluted} response matrix and actual count rate, the $\nu F_{\nu}$ spectrum, and the corner plot of the posterior parameters for the photon spectrum model. The model in the left panel is Band*$highecut$, and the model in the right panel is (Band+$powerlaw$)*$highecut$.}}
    \label{fig:prompt_spec}
\end{figure*}

\begin{figure}[htp]
    \centering
    \includegraphics[width=0.7\textwidth]{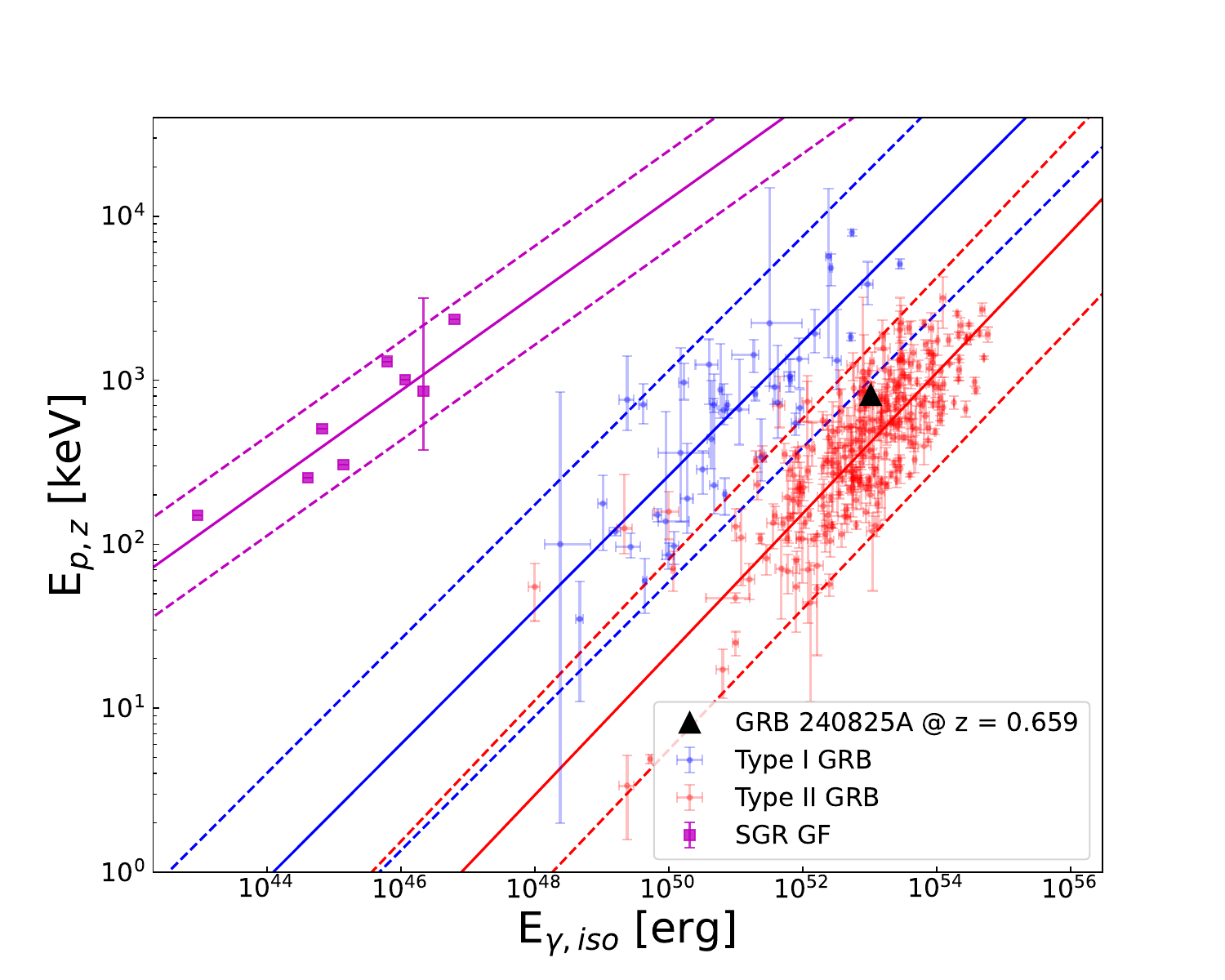}
    \caption{The $E_{\rm p,z}$–$E_{\gamma,\rm iso}$ diagram. The light blue and light red points represent the data of Type I and Type II GRB, and Soft Gamma-ray Repeater Giant Flare (SGR GF), respectively, with known distances \citep{2006Natur.444.1010Z,2009ApJ...703.1696Z,minaev2020p,2020AstL...46..573M}. The corresponding dashed lines indicate the 2$\sigma_{\rm cor}$ confidence regions of the correlations. The black triangle denotes GRB 240825A.}
    \label{fig:amati}
\end{figure}

The light curves of prompt emission detected by different instruments in different energy bands are shown in Figure \ref{fig:prompt_LC}. \wyAdd{Based on the $T_{90}$ ($T_0$ + [1.15, 5.12] s) from GBM \citep{2024GCN.37273....1F}, we performed a time-integrated spectral analysis for this interval (blue dashed line) with {Swift-BAT}, Fermi-GBM and LAT joint-fit.} \wyAdd{In the spectral fitting process, the first photon spectral model we consider is the Band function \citep{1993ApJ...413..281B} with a high-energy cutoff (i.e., Band*$highecut$), which is the Band function multiplied by $e^{-E/E_f}$.}
where $E_f$ is the characteristic parameters related to the high-energy spectral cutoff, with units of MeV. The second photon spectral model includes an additional power-law ($N(E) \propto E^{-\hat{\gamma}}$) component that extends into the high-energy range, expressed as ``(Band + $powerlaw$)*$highecut$'', \wyAdd{with the energy peak of $(2 - \hat{\gamma}) E_f$}. For the two-photon spectral models mentioned above, we applied Bayesian inference \citep{2019PASA...36...10T,van2021bayesian} for model comparison and parameter estimation. The likelihood functions for different instruments are defined as follows: the statistics used for Swift-BAT is $\chi^2$, and the statistic for Fermi data is {\tt pgstat}, refer to the {\tt XSPEC} manual\footnote{\url{https:// heasarc.gsfc.nasa.gov/xanadu/xspec/manual/XSappendixStatistics.html}}. For Bayesian inference, we adopt the {\tt PyMultiNest} sampler \citep{2014A&A...564A.125B}. The results are shown in Figure \ref{fig:prompt_spec}, with the upper and middle panels presenting the photon count spectrum and the $\nu F_{\nu}$ spectrum, respectively, both obtained by folding the response function with the best-fit parameters. The bottom panel in Figure \ref{fig:prompt_spec} shows the posterior parameter distributions. For model comparison, \wyAdd{we use} the logarithm of the Bayes factor, written as
\begin{equation}
    \log \rm{BF}^{A}_{B} \equiv \log (\mathcal{Z}_{A}) - \log (\mathcal{Z}_{B}),
\end{equation}
where $\log (\mathcal{Z}_{\rm A})$ and $\log (\mathcal{Z}_{\rm B})$ are the evidences from two different models, respectively.
Here, the model with the additional power-law component achieved higher evidence, and the Bayesian factor $\log$BF showed 83 indicating that the data provide stronger support for this model \citep{Thrane2019,jeffreys1998theory}. \wyAdd{The energy peak in $\nu F_{\nu}$ spectrum of the Band component is $491.65_{-7.94}^{+7.81}$ keV, with a low-energy spectral index $\alpha$ = $-0.62_{-0.02}^{+0.02}$ and a high-energy spectral index $\beta$ = $-2.55_{-0.05}^{+0.05}$. The spectral index of the second powerlaw component is $\hat{\gamma}$ = $1.64_{-0.01}^{+0.02}$, and the high-energy spectral cutoff in $\nu F_{\nu}$ spectrum is $76.03_{-6.19}^{+7.46}$ MeV. }
\wyAdd{The fluence observed in the $\gamma$-ray band (1–10,000 keV) is $S_\gamma =3.74_{-0.12}^{+0.16} \times 10^{-5}~\rm{erg/cm^{2}}$.} Based on 
\begin{equation}
    E_{\gamma,\text{iso}} = \frac{4 \pi d_L^2 k S_{\gamma}}{1+z},
    \label{eq:E_gamma_iso}
\end{equation}
where $d_L$ is the luminosity distance, $k \equiv \int^{10^4/(1+z)}_{1/(1+z)}EN(E)dE/\int^{e2}_{e1}EN(E)dE$ ($e1$ and $e2$ are the energy bands of the detector) is the correction factor \citep{2001AJ....121.2879B}, the isotropic $\gamma$-ray energy $E_{\gamma,\text{iso}}$ is 
{$1.72_{-0.06}^{+0.07} \times 10^{53}~\rm{erg}$.} This result is consistent with the report from Konus-Wind \citep{2024GCN.37302....1F} and falls within the range of Type II GRBs \citep{2006Natur.444.1010Z,2009ApJ...703.1696Z} in the Amati relation \citep{amati2002intrinsic}, as shown in Figure \ref{fig:amati}. In addition, we noticed that the other two papers provided a more detailed analysis and discussion of the prompt emission \citep{2025ApJ...984L..45Z,2025ApJ...985L..30W}.

\subsubsection{Extended high-energy observations}
\label{sec:he_af}
Swift-XRT and Fermi-LAT detected emission with extended duration from this source, ranging from X-rays to $\gamma$-rays \citep{2024GCN.37294....1G,2024GCN.37288....1D}. We reprocessed the long-duration Swift-XRT data, selecting grades 0–12 for the Photon Counting (PC) mode and grades 0–2 for the Windowed Timing (WT) mode, and restricted the energy range to 0.3–10 keV. The light curves were rebinned using the step-binning method implemented in \texttt{xrtgrblc}. Time-resolved spectral information was taken into account when converting the count rates to energy fluxes, based on the spectral evolution data provided by the online burst analyzer tool\footnote{\url{https://www.swift.ac.uk/burst\_analyser/}} 
(see more details in \citealp{2007A&A...469..379E,2009MNRAS.397.1177E}). For the Fermi-LAT data reduction, we selected photon events within the energy range of 100 MeV to 100 GeV and the 0–2200 s interval, using the {\tt TRANSIENT} event class and the {\tt FRONT+BACK} type. To minimize contamination from the Earth’s limb, photons with zenith angles exceeding 100$^{\circ}$ were excluded. Subsequently, we identified good time intervals by applying the quality filter condition ({\tt (DATA\_QUAL$>$1 \&\& LAT\_CONFIG==1)}). In the standard unbinned likelihood analysis procedure\footnote{\url{ https://fermi.gsfc.nasa.gov/ssc/data/analysis/scitools/}}, we selected a 10$^{\circ}$ region centered on the location of GRB 240825A as the region of interest (ROI). The initial model for the ROI region, generated by the {\tt make4FGLxml.py} script\footnote{\url{http://fermi.gsfc.nasa.gov/ssc/data/analysis/user/}}, includes the galactic diffuse emission template ({\tt gll\_iem\_v07.fits}), the isotropic diffuse spectral model for the {\tt TRANSIENT} data ({\tt iso\_P8R3\_TRANSIENT020\_V3\_v1.txt}) and all the Fourth Fermi-LAT source catalog~\citep[{\tt gll\_psc\_v31.fit};][]{2020AbdollahiApJS} sources. We defined GRB~240825A as a point source in the model file, with the model set to {\tt PowerLaw2}\footnote{\url{https://fermi.gsfc.nasa.gov/ssc/data/analysis/scitools/source_models.html}}, \wyAdd{which is simple power law with an integral number of counts between two energies as the normalization}. An automatic rebinning algorithm was employed, ensuring at least 10 photons per bin and TS $>$ 16 to guarantee that the flux in each bin is significant. \wyAdd{The results of the data analysis in this section are used for the multiband afterglow fitting in Section\ref{sec:af}}.

\subsection{Optical Observations}
\label{sec:OO}
\subsubsection{SVOM/C-GFT}
The Chinese Ground Follow-up Telescope \citep[C-GFT,][]{2016arXiv161006892W} is a dedicated ground-based instrument for the SVOM mission, located at the Jilin Station of the Changchun Observatory, NAOC (National Astronomical Observatories, Chinese Academy of Sciences). Its coordinates are $126^{\circ}19'49.7''$ East, $43^{\circ}49'27.8''$ North, with an elevation of 320 meters. Detailed information on the C-GFT will be presented in Wu et al. (in preparation).
The C-GFT collaborates closely with the French-Mexican Ground Follow-up Telescope (FM-GFT, also known as the Colibri Telescope) located in Mexico \citep{2022SPIE12182E..1SB}. Together, they form a critical component of a global telescope network designed to provide rapid responses to SVOM alerts. Additionally, they are integrated with the Ground-based Wide Angle Camera array (GWAC) \citep{2021PASP..133f5001H, 2023NatAs...7..724X}, 
enabling comprehensive monitoring of gamma-ray burst (GRB) evolution across all phases, from the prompt emission to late afterglows, in the optical band (and in the near-infrared for FM-GFT).

C-GFT is equipped with two focal-plane instruments \citep{2022RAA....22e5009N}, the Camera with Three CHannels  (CATCH) and LArge-FOV Transient Imager with CMOS (LATIOS), which can be switched using the secondary mirror during daytime. CATCH is mounted at the Cassegrain focus, having an effective FOV of $21^\prime\times21^\prime$, and capable of imaging simultaneously in the SDSS $g$, $r$, and $i$ bands. LATIOS is a CMOS camera mounted at the prime focus. It currently uses a sCMOS imaging sensor (Balor F17-12, $4k \times 4k, 12\ \mu$m) 
to achieve an effective field of view (FOV) of $1.28^{\circ} \times 1.28^{\circ}$, with a pixel scale of $1.13$ arcsec/pixel. A filter slider mechanism enables the camera to do multi-color observations in the SDSS photometric system, containing a set of high-transmission SDSS standard filters of all dielectric coating technology \footnote{https://www.asahi-spectra.com/opticalfilters/sdss-d.asp}. During the commissioning phase, C-GFT typically operated in the prime focus mode in order to get a wide sky coverage.

When a new GRB is detected by SVOM or other satellites, C-GFT automatically responds to the GRB trigger alert, transmitted from the GRB alert network, within seconds. Upon receiving an alert, the control system instantly generates an observation plan containing new pointing instructions, which can interrupt the current observation sequence. If the GRB target is accessible to C-GFT at that moment, the telescope will be re-pointed at it and follow-up observations be started in no less than one minute after receiving the alert message, thanks to an all-sky dome and an altazimuth mount. Otherwise, it will either be scheduled on a waiting list or just be rejected. 

\begin{figure}[htp]
    \centering
    \includegraphics[width=0.7\textwidth]{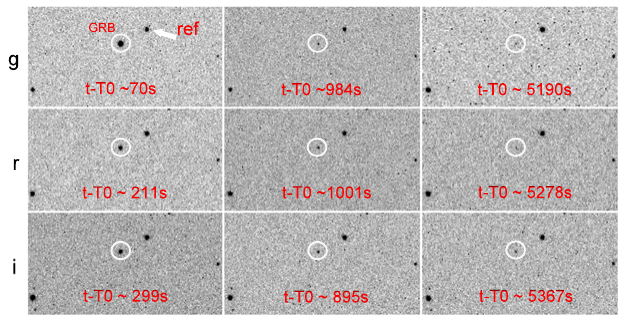}
    \caption{A sample of C-GFT LATIOS images obtained for GRB 240825A at the early, middle, and late time (from left to right) using the SDSS $g$ (top), $r$ (middle), and $i$ (bottom) filters. All images are aligned (North up, East left) with a uniform scale of 280 arcsec $\times$ 230 arcsec. The GRB afterglow is marked with a white circle, and the reference star is marked with an arrow. }
   \label{fig:cgft-image}
\end{figure}

The first C-GFT image of GRB 240825A was taken with LATIOS starting from 15:54:05 UT on August 25, 2024, 
which was only 66 seconds after the BAT trigger. Continuous LATIOS follow-up observations lasted for about 89 minutes in total. As a result, a sequence of images of exposure time of 10 sec for each have been obtained in the SDSS $g$, $r$, and $i$  bands. Their respective exposure mid-time parameters and GRB photometric results are listed in Table~\ref{tab_cgft}.
Our data reduction was carried out as follows. After bias subtraction and dark correction, astrometric calibration was performed using astrometry.net \citep{2010AJ....139.1782L}. Differential aperture photometry was then conducted using the IRAF photometric package. Due to the absence of flat-field data from the same night, only the nearest bright star was selected as the photometric reference. This reference star was validated through flux stability cross-checking with other bright stars in the vicinity of the GRB target. The coordinates of this reference star are RA(J2000)=22h58m14.76s and Dec(J2000)=+01d01$'57.5''$. For flux calibration, we utilized the Pan-STARRS Release 1 \citep[PS1:][]{2016arXiv161205560C} catalog available in VizieR \citep{vizier2000}.
Both the reference star and GRB target are denoted in Figure \ref{fig:cgft-image}. Data binning (as denoted by ``$*$'' in Table~\ref{tab_cgft}) or image stacking (as denoted by ``${\blacktriangle}$'' in Table~\ref{tab_cgft})  was employed if necessary to mitigate the effects of outliers or enhance the signal-to-noise ratios, especially for the late time when the afterglow was faint.

\subsubsection{SVOM/VT}

The VT (Visible telescope) is an optical telescope onboard the Space-based multiband Variable Objects Monitor \citep[SVOM,][]{2016arXiv161006892W} mission. The effective aperture is 43 cm. The FOV is 26$\times$26 square arcminute, giving the pixel scale of 0.76 arcseconds. 
It conducts the observation with two channels VT\_B and VT\_R simultaneously, covering the wavelength of 400-650 nm and 650-1000 nm respectively.
Detailed information on VT could be referred to Qiu et al. (2025)(in preparation). 
During the commission phase, GRB 240825A was observed by VT three epochs via Target Of Opportunity (ToO) mode on
August 26, 2024, September 8, 2024 and September 23, 2024.
The exposure time was 20 seconds per frame, with a total of 405 frames acquired for VT\_B and 434 frames for VT\_R band, corresponding to about $\sim140$ minutes of observation time.
All the data were processed in a standard manner using the IRAF\footnote{IRAF is distributed by the National Optical Astronomical Observatories, which are operated by the Association of Universities for Research in Astronomy, Inc., under cooperative agreement with the National Science Foundation.} package., including zero correction, dark correction, and flat-field correction. 
After the pre-processing, the images for each band obtained during each epoch were stacked to increase the signal-to-noise ratio. The counterpart was clearly detected in all stacked images.
Aperture photometry was adopted. The aperture radius and full aperture radius set to be 1.5 pixels and 7.0 pixels, respectively. For each image aperture correction was done to derive the final brightness. The photometric results are listed in Table \ref{tab_vt}, in which the magnitudes were calibrated in AB magnitude and not corrected for the Galactic extinction.

\subsubsection{Swift-UVOT}
{The data reduction for early optical observations with Swift-UVOT, based on observations in the $V$, $B$, $U$, $W1$, $M2$, $W2$, and $white$ bands over several epochs, was performed using {\tt HEAsoft}\footnote{\url{https://heasarc.gsfc.nasa.gov/ftools}}, version 6.32.1.} We only took data in event mode and started with Level 1 UVOT products. For images without detections, upper limits were set by assuming that count rates would have reached a signal-to-noise ratio of 3. Since only upper limits were detected in W1, M2 and W2 bands, reliable detections were obtained in the V, U, and white bands. The photon count rates were measured using 5 arcsec apertures in standard aperture photometry, and the background was measured in a nearby region without sources. For images without detections, upper limits were set by assuming that count rates would have reached a signal-to-noise ratio of 3. All photometric data are tabulated in Table \ref{tab_uvot}. It is important to note that all white-band exposures, as well as the last three exposures of the $U$-band are affected by the Small Scale Sensitivity (SSS) effect\footnote{\url{https://swift.gsfc.nasa.gov/analysis/uvot_digest/sss_check.html}}.

\subsubsection{HMT/T80/PAT17/ALT100C}

The Half-Meter-Telescope (HMT), an unfiltered instrument located in Nanshan, Xinjiang, China, started observations at 16:06:38 UT on August 25, 2024, corresponding to 819 seconds post-trigger. The optical emission was clearly detected in the initial frame with unfiltered magnitude $m_{c} \sim 17.0$. Concurrent observations of the afterglow on the first night were conducted utilizing an 800 mm unfiltered telescope (T80) and a 17-inch photometric auxiliary telescope (PAT17), both stationed at Nanshan, Xinjiang. Subsequent monitoring of the afterglow was carried out using the 100-cm C telescope (ALT100C) of the JinShan project, located in Altay, Xinjiang, China. Photometric calibration was performed using the Gaia-DR3 \citep{2023A&A...674A...1G} catalog for the clear band and the Pan-STARRS DR2 catalog for other bands. Photometric data are tabulated in Table \ref{tab_xm} (Gaia-calibrated) \& Table \ref{tab_xm2} (Pan-STARRS-calibrated), respectively.

\subsubsection{LCOGT}
The Las Cumbres Observatory Global Telescope Network \citep[LCOGT;][]{2013PASP..125.1031B} observed GRB 240825A using the 1 m telescope. Observations of the afterglow of GRB 240825A began at 01:41:29 UT on August 26, 2024, $\sim$9.81 hr after the GBM trigger. A total of 10 observing runs were conducted between August 26 and 31, resulting in a set of B, V, R, and I band images. 
The exposure time of each image was set to 300 s, based on the trend of brightness estimated from other observations' measurements and the limiting magnitude of the LCOGT telescopes.
The photometry calibration is carried out using the USNO-B1 catalogs \citep{2003AJ....125..984M}. The source is clearly detected by stacking the images from the first nine observations. All photometric data are tabulated in Table~\ref{tab_LCOGT}.

\subsubsection{NOT}
We triggered the 2.56 m Nordic Optical Telescope (NOT) at the Roque de los Muchachos Observatory, La Palma, Spain, several times, utilizing the Alhambra Faint Object Spectrograph and Camera (ALFOSC). Photometric calibration was performed using the Pan-STARRS DR2 catalog \citep{2018AAS...23143601F}. Photometric data are tabulated in Table \ref{tab_NOTVLT}.

\subsubsection{GTC}
\begin{figure*}[htp]
    \centering
    \includegraphics[width=\textwidth]{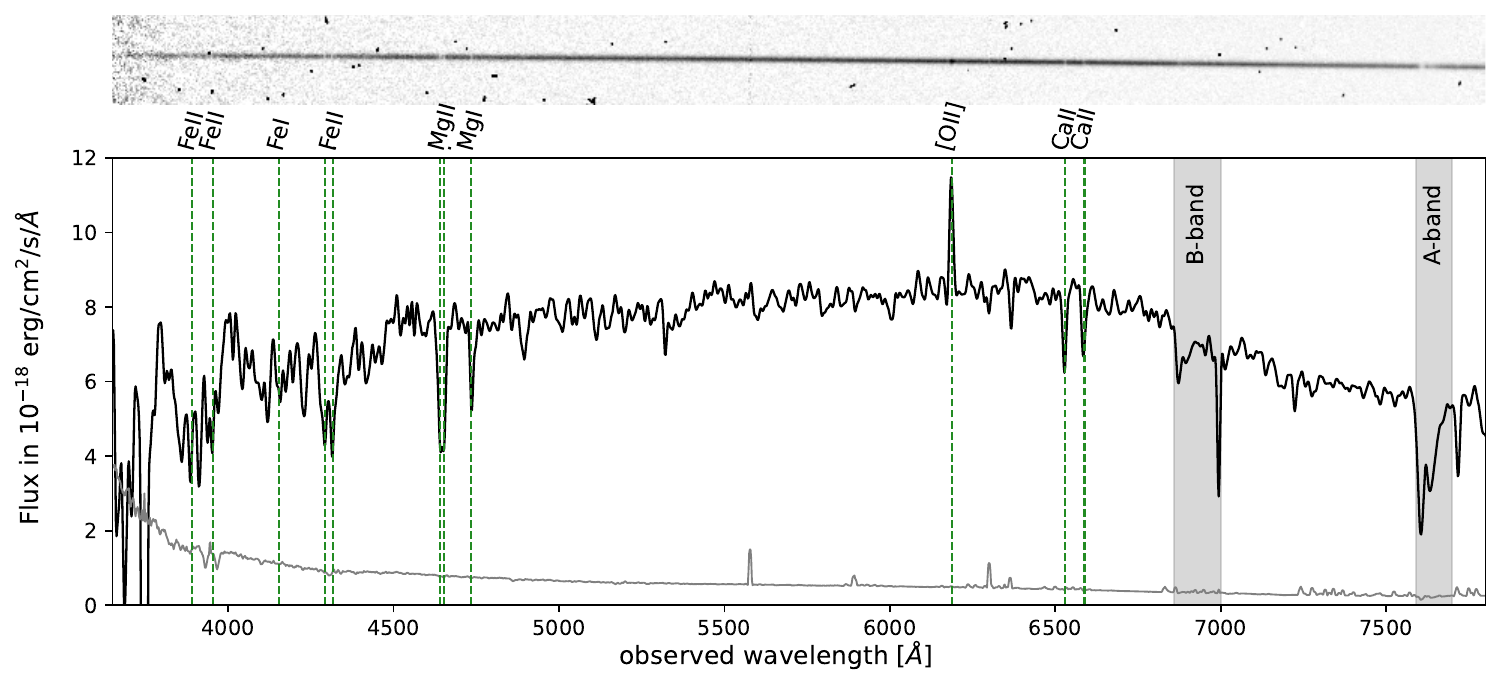}
    \includegraphics[width=\textwidth]{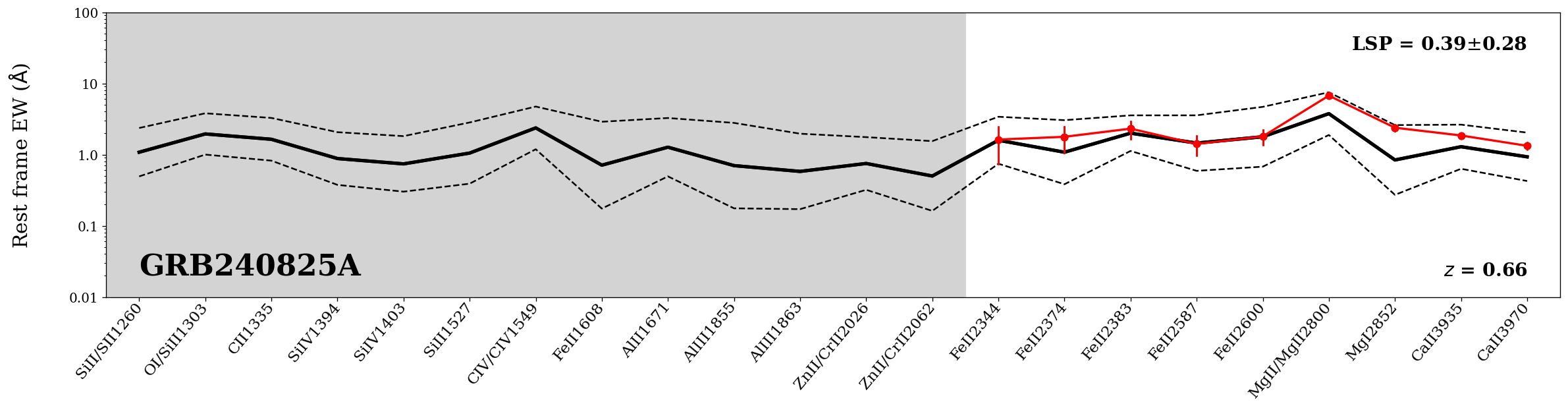}
    \caption{Top: GTC spectrum of the GRB afterglow, showing multiple absorption features and emission due to the [\oii] doublet. Bottom: Line Strength Diagram, comparing the strength of the lines covered by the GTC spectrum (in red) with those of the sample published by \citet{Fynbo2009} following the method of \citet{deUgartePostigo2012}. The shaded grey features are those nor covered by our spectral range.}
   \label{fig:gtcspec}
\end{figure*}

The 10.4 m Gran Telescopio CANARIAS (GTC) telescope, at the Roque de los Muchachos Observatory, on the Island of La Palma (Spain) observed the afterglow of GRB\,240825A. The observation started on August 26 at 02:04:10 UT, 10.190 hr after the burst onset. It was performed with OSIRIS+ \citep{Cepa2000} and consisted of $3\times30$ s acquisition images in $r$-band followed by $3\times900$ s spectra using grism R1000B, covering the range between 3600 and 7800 {\AA}, with a slit of 1$^{\prime\prime}$ aligned with the parallactic angle, providing a resolving power of R $\sim 600$. The mean epoch of the spectroscopic observation was 02:27:08 UT, 10.569 hr after the burst. The spectrum displays a clear continuum with absorption features due to \feii, \mgii, \mgi, and \caii\ and emission due to [\oii] at a common redshift of 0.6596$\pm$0.0006. The spectrum is shown in the top panel of Figure \ref{fig:gtcspec}.

Table~\ref{tab_gtc} displays the equivalent width measurements of the prominent absorption features measured in the GTC spectrum. These values are used to calculate the line strength parameter (LSP), which allows us to compare the strength of the absorption features of a GRB spectrum with those of a general GRB line-of-sight sample, independently of the spectral resolution or wavelength coverage. Following the prescription of \citet{deUgartePostigo2012}, we obtain a value of LSP $= 0.39\pm0.29$, which implies that the lines in this spectrum are slightly stronger than those of the typical GRB sight-line, corresponding to the 72 percentile of the samples of \citet{Fynbo2009} and \citet{deUgartePostigo2012} (the lines are stronger than those of 72\% of the sample). The bottom panel of Figure \ref{fig:gtcspec} shows the Line Strength Diagram, which shows the comparison with the sample in more detail, comparing individual features. The line in black represents the average strength of the lines in the sample, and the dotted lines mark the standard deviation from that value. The red dots are the measurements from our spectrum. In this case, the $\feii$ lines are almost equal to the sample, whereas the $\mgii$ and $\mgi$ features are 1-$\sigma$ stronger than the sample. The $\caii$ lines are slightly stronger than those of the sample.

Follow-up imaging observations were obtained using the OSIRIS+ in $r$ and $z$ bands and with the HiPERCAM instrument \citep{Dhillon2021}, which simultaneously observes in all Sloan filters $u$, $g$, $r$, $i$ and $z$. We performed aperture photometry using SDSS field stars as a reference. The results are presented in Table~\ref{tab_NOTVLT}.

\begin{table}
\begin{center}
\caption[]{ Equivalent Widths of Absorption Features in the GTC Spectrum.}\label{tab_gtc}
 \begin{tabular}{rcc}
  \hline\noalign{\smallskip}
Feature       & Wavelength ({\AA}) & EW ({\AA})   \\   
  \hline\noalign{\smallskip}
FeII$\lambda$2344        	 	& 3886.3    & (2.70$\pm$1.51)$^\ast$ \\
FeII$\lambda$2374	       		& 3937.5    & (2.95$\pm$1.23)$^\ast$ \\
FeII$\lambda$2383	       		& 3952.7    & 3.84$\pm$1.15 \\
FeII$\lambda$2587 	 	& 4293.6    & 2.34$\pm$0.75 \\
FeII$\lambda$2600 	 	& 4316.5    & 2.99$\pm$0.77 \\
MgII$\lambda$2800 blend	& 4647.4    & 11.17$\pm$0.69 \\
MgI$\lambda$2852 	       	& 4737.0    & 3.96$\pm$0.47 \\
CaII$\lambda$3935	       		& 6529.5    & 3.08$\pm$0.26  \\
CaII$\lambda$3970 	      	& 6586.5    & 2.20$\pm$0.27  \\
  \noalign{\smallskip}\hline
\end{tabular}
\tablecomments{0.86\textwidth}{``${\ast}$" indicates formally a non detection.}
\end{center}
\end{table}

\subsubsection{VLT}
We also carried out observations of the optical counterpart using the X-shooter instrument \citep{vernet_x-shooter_2011}, a multiwavelength medium-resolution spectrograph installed at the European Southern Observatory (ESO) Very Large Telescope (VLT) UT3 (Melipal) under the program 110.24CF (PI: N.~Tanvir). 
Our sequence started on 2024 August 26 at 02:18:03 (10.42 hr after the trigger) with photometric measurements of the optical counterpart in g$^{\prime}$, r$^{\prime}$, and z$^{\prime}$ SDSS filters using the acquisition and guiding camera of the X-shooter (see Table \ref{tab_NOTVLT}). The spectroscopic observations were then obtained at a mid-time of 11.35~hr after the {\it {Swift}} trigger. Our spectra cover the wavelength range $3000 - 21000$\AA, and consisted of four exposures, each 1200 s. The full data set was reduced using the ESO pipeline following the strategy detailed in \cite{Goldoni2006} and \cite{Modigliani2010}. Wavelengths were corrected to the vacuum-heliocentric system.

In the calibrated spectrum, a trace due to the afterglow emission is well detected over the entire wavelength range and several absorption features were identified due to \feii, \feii{}{*}, \mnii, \mgii, \mgi, \caii, \cai\, and \nai. Using all these metal absorption lines, we measured $z_{abs}=0.6593\pm0.0001$ as the redshift of the GRB \citep{GCN_xshooter}.
In addition, at $z_{em}=0.6591\pm0.0001$, we also detect multiple bright emission lines produced by the GRB host galaxy as the [O II] and [O III] doublets, H$\alpha$, and H$\beta$, together with [N II] and [S II] doublets (partially affected by sky lines). We note that the small shift between the redshift measured from the emission and absorption lines is consistent within the 1$\sigma$ errors and corresponds to a velocity shift of $\sim$$-36\,$km~s$^{-1}$ towards the blue of the emission lines with respect to the absorption lines. The properties of the host galaxy will be investigated in more detail in a future work (Schneider et al., in preparation).

\subsubsection{Asiago/REM/TNG/LBT}
We obtained an optical Sloan-$i$ band observation with the 0.67/0.92\,m robotic Schmidt telescope, located on Mount Ekar at the Asiago Observatory (Italy). The observation started on 2024 August 25 at 21:39:05 UT, which is $\sim$ 5.8 hr after the GRB trigger and lasted for 20 minutes. The data were reduced using the Snoopy pipeline\footnote{{\sl Snoopy} is a package for SN photometry using point-spread function (PSF) fitting and/or template subtraction developed by E. Cappellaro. A package description can be found at \url{http://sngroup.oapd.inaf.it/snoopy.html}.}, which was used also to perform PSF photometry, calibrated against the Pan-STARRS catalog. The astrometry was calibrated against field stars in the Gaia DR3 catalog.
The measured magnitudes are listed in Table \ref{tab_REM}.

We obtained optical/near-infrared (NIR) observations with the 0.6\,m robotic Rapid Eye Mount telescope \citep[REM,][]{Zerbi+01,Covino+04}, located at the ESO at La Silla (Chile). The observations started on  2024 August 26 at 00:28:43 UT, which is $\sim$ 8.6 hr after the GRB trigger and lasted for about 1 hour. Data reduction was performed automatically by the REM reduction pipeline: after bias subtraction, non-uniformities were corrected using a normalized flat-field frame processed with tools from the Swift Reduction Package (SRP)\footnote{\url{http://www.me.oa-brera.inaf.it/utenti/covino/usermanual.html}}. NIR data were sky-subtracted using the median of individual frames. Frame registration was performed using the Python-based software Astroalign \citep{Beroiz19}, and astrometric solutions were derived against Gaia DR3 stars \citep{GAIA}. An uncataloged source was detected, at a position consistent with the optical afterglow, in the stacked $r$- and $J$-band images. Aperture photometry was performed on the candidate with the SExtractor package \citep[Version 2.28.1][]{SExtractor} and results were calibrated against Pan-STARRS DR2 \citep{PS1} and 2MASS \citep{2MASS} sources for the optical and NIR images, respectively. 

\begin{figure}
    \centering
    \includegraphics[width=0.7\textwidth]{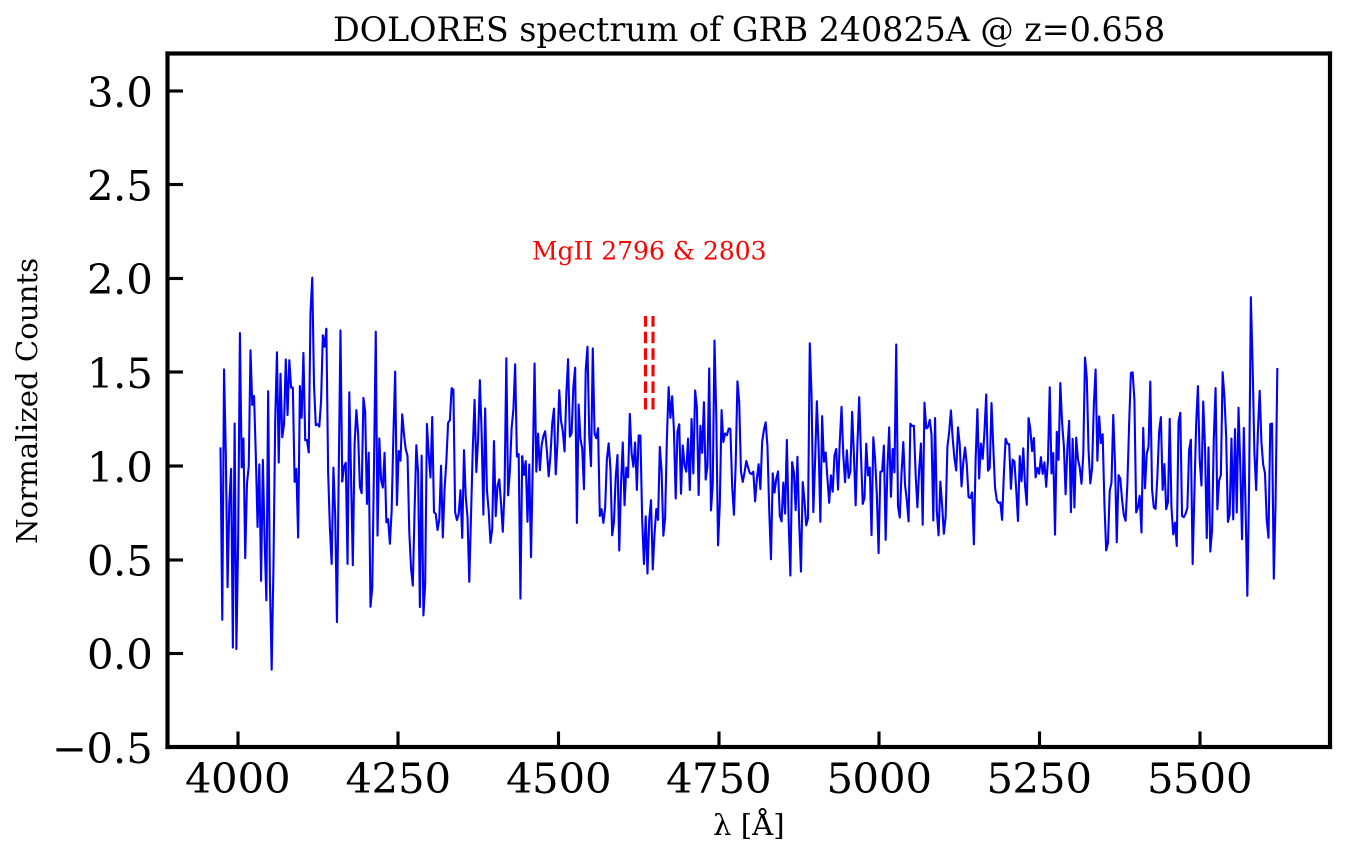}
    \caption{DOLoRes spectrum of GRB 240825A.}
    \label{fig:spec_TNG}
\end{figure}

The field of GRB\,240825A was also observed with the DOLoRes camera in spectroscopic mode mounted on the 3.6\,m Telescopio Nazionale Galileo (TNG) at La Palma, Spain. Observations were carried out with the LR-B grism, covering the range 3500-8000 \AA, with $1''$ slit. The observations consisted of a spectrum carried out at a mean time of 7.4 hr after the burst. The optical afterglow is detected in the acquisition image, and aperture photometry was carried out with the \texttt{SExtractor} package and calibrated against the Pan-STARRS DR2 catalog. The spectrum was reduced following standard procedure with the IRAF package \citep[Version 2.16.1][]{IRAF}, including bias subtraction, flat-field correction, wavelength calibration, and extraction. Despite the modest SNR we identified a single absorption feature consistent with Mg II at a redshift of $z=0.658$ (Figure \ref{fig:spec_TNG}), in agreement with the results reported by \cite{GCN_xshooter}. Further observations were performed with the Near Infrared Camera Spectrometer (NICS) mounted on the TNG telescope \citep{TNG_NICS} at a mid-time $t-t_0 = 2.4$\,d. Data were reduced and stacked together with the \texttt{jitter} tool of the \texttt{eclipse} package, and the NIR afterglow was clearly detected. Aperture photometry was performed with SExtractor and calibrated against the 2MASS catalog. We also obtained $z$-band observations with the DOLoRes camera mounted on the TNG at a mid-time of about 13.4 days from the GRB trigger. The images were reduced and sky-subtracted with the  \texttt{jitter} tool of the \texttt{eclipse} package. The source was visible at the position of the optical afterglow, aperture photometry was performed with the SExtractor package and calibration was against the Pan-STARRS DR2 catalog.
All measured magnitudes are listed in Table \ref{tab_REM}.

On 2024 November 9, we also imaged the host in the $ r^\prime z^\prime$-bands with the Large Binocular Cameras \citep[LBCs; ][]{Giallongo2008a} mounted on the Large Binocular Telescope (LBT) on Mt. Graham, Arizona, USA. The observations started at 03:58 UT, for a total of 3000 s in each band. LBT imaging data were reduced using the dedicated data reduction pipeline \citep{Fontana2014a}. The astrometry was calibrated against field stars in the Gaia DR3 catalog. We performed aperture photometry using \texttt{DAOPHOT} and \texttt{APPHOT} under \texttt{PyRAF/IRAF} and results were calibrated against the Pan-STARRS survey. All measured magnitudes are listed in Table \ref{tab_REM}.

\subsubsection{MISTRAL}
We finally carried out additional observations of the optical afterglow using the MISTRAL instrument mounted on the 193 cm telescope at Observatoire de Haute-Provence (OHP, France). The mid-time of these observations was  2024 August 25, 22:40 UT, corresponding to 6.85 hr after the GRB trigger. Three exposures of 5 minutes were obtained in the SDSS r$'$ band, leading to a total exposure time of 15 minutes. Data were reduced using standard bias- and flat-removal procedures, and the photometry was calibrated  against nearby stars from the Pan-STARRS catalog. The magnitude we derived is listed in Table \ref{tab_REM}.

\section{Physical Implications for Afterglow} \label{sec:af}
\subsection{Afterglow model}
The interaction between the relativistic outflow of the GRB and the circumburst medium can drive a pair of relativistic shocks: an FS and an RS. Electrons accelerated by these shocks produce multiwavelength emission through synchrotron radiation and inverse Compton scattering. \wyAdd{We will analyze the afterglow data in the NIR, optical, X-ray, and GeV bands given in Sections \ref{sec:he_af} and \ref{sec:OO}, as well as the photometric data from Mephisto \citep{2025ApJ...979...38C}.}

To derive the temporal profile of the afterglow light curve, we fitted the $g$-band data, which corresponds to the filter used by C-GFT for its initial observations, using a broken power-law (BPL) function, given by:
\begin{equation}
    F=F_0 \Bigg[ \bigg(\frac{t}{t_b}\bigg)^{\alpha_{g,1}\omega}+ \bigg(\frac{t}{t_b}\bigg)^{\alpha_{g,2}\omega} {\bigg]}^{-1/\omega}.
\end{equation}
The term $F_0$ represents the normalization constant, $t_b$ is the break time, $\alpha_{g,1}$ and $\alpha_{g,2}$ are the indices of the two power-law segments in g-band, and $\omega$ represents the sharpness of the break. Fitting with the BPL function is a simple and efficient method to characterize the afterglow properties \citep[e.g.][]{2008ApJ...675..528L,2015ApJS..219....9W}. We adopted a Gaussian likelihood function, assuming independent normal errors with known variances for each data point. From the fit, we obtained the break time $t_b\sim255$ s, with $\alpha_{g,1}=1.82_{-0.04}^{+0.07}$ and $\alpha_{g,2}=0.98_{-0.02}^{+0.02}$ for the indices of the two power-law segments, and $\omega$ = -3.58$^{+1.01}_{-0.91}$. This slope transition can be interpreted as a shift from RS domination to FS domination, corresponding to one of the ``flattening cases" of the RS and FS combinations discussed in previous works \citep{2003ApJ...595..950Z,2007MNRAS.378.1043J,2015ApJ...810..160G}. Based on the current fit of $\alpha_{g,1}$ and the previous work by \cite{2015ApJ...810..160G}, we can approximately obtain the electron spectral index of the RS as $p_r \sim 2.1$ in the thin shell scenario for II Case, where $F_{\nu} \propto t^{-(27p+7)/35}$. Adopting a thick shell scenario, in which the flux evolves as $F_{\nu} \propto t^{-(73p+21)/96}$, the predicted $p_r$ is essentially the same as in the thin shell scenario. Thus, in the absence of earlier-time observations, we cannot distinguish whether the RS crossed a thick or thin shell. Based on $\alpha_{g,2}$ and the closure relations in the interstellar medium (ISM) case \citep{2004IJMPA..19.2385Z,2006ApJ...642..354Z}, for the slow cooling scenario ($\nu_m < \nu_g < \nu_c$, where $\nu_m$ is the characteristic synchrotron frequency of the electrons at the minimum injection energy, $\nu_c$ is the cooling frequency), with %
$F_{\nu} \propto t^{-3(p-1)/4}$, the electron spectral index of the FS can be predicted to be $p_f \sim$ 2.3. For the slow cooling scenario in a wind environment, the decay index $\alpha$ follows the same closure relation. Therefore, based on the decay slope alone, it is difficult to distinguish whether the environment is ISM or wind. Thus, in the following model fitting, we only consider the RS in the thin shell scenario within an ISM environment, which does not imply that we favor this model, but rather is the simplest model based on the available observations.

For afterglow modeling, numerous paradigms for numerical calculations have been established, generally starting from shock dynamical evolution \citep{1999MNRAS.309..513H,2013MNRAS.433.2107N,Zhang_2018pgrb.book.....Z}, solving the electron continuity equation \citep{Sari_1998,2001ApJ...548..787S,2006MNRAS.369..197F,2008MNRAS.384.1483F,2012MNRAS.427L..40K}, calculating the synchrotron radiation \citep{Sari_1998,1999ApJ...517L.109S} and synchrotron self-Compton \citep[SSC; e.g.][]{2008MNRAS.384.1483F,2009ApJ...703..675N,2018ApJS..234....3G} scattering of electrons, and considering the impacts of various effects, including Klein-Nishina \citep[KN; e.g.][]{2008MNRAS.384.1483F,2009ApJ...703..675N} effects,  $\gamma\gamma$ annihilation effects \citep[e.g.][]{1967PhRv..155.1408G,2011ApJ...732...77M,2018ApJS..234....3G,2022ApJ...931..150H}, synchrotron self-absorption \citep[SSA; e.g.][]{1999ApJ...527..236G}, as well as geometric and observational effects \citep{1997ApJ...485L...5W,2018ApJS..234....3G}.
Here, we use the Python-wrapped Fortran package {\tt ASGARD} to perform the numerical calculations of the FS radiation under the various effects mentioned above \citep{2024ApJ...962..115R}.
The calculation of the RS contribution in the early afterglow follows the analytical method described in \cite{2013ApJ...776..120Y}, which includes only synchrotron emission. Given the large uncertainties and limited number of data points in the GeV energy band, we neglect the SSC contribution. This simplification may to some extent affect the estimation of the characteristic frequencies in the synchrotron emission.
In this work, we assume that the jet has a “top-hat” structure. Therefore, under this model, the flux density at a certain time and frequency is:
\begin{equation}
\begin{aligned}
    F_{t, \nu} = & \, F(t,\nu, \Gamma_0,\epsilon_{e,f},\epsilon_{B,f},\theta_j,E_{\rm k,iso},p_{f},  \epsilon_{e,r},\epsilon_{B,r},p_{r},n_0)  + F_{\rm host}(\nu),
\end{aligned}
\label{eq:fs_rs}
\end{equation}
where $\Gamma_0$ is the initial Lorentz factor, $\epsilon_e$($\epsilon_B$) is the fraction of the shock energy density converted into energy of relativistic electrons (magnetic field), $\theta_j$ is the half-opening angle in radians, $E_{\rm k,iso}$ is the isotropic kinetic energy, $p$ is the electron energy distribution index, and $n_0$ is the medium number density. The additional subscripts $f$ and $r$ represent the FS and RS, respectively. Additionally, the host galaxy contribution $F_{\nu}$ is derived from the SED fitting results presented by \cite{2025ApJ...979...38C}. The model also accounts for several additional factors. For example, for the optical data, it adopts the \cite{1999PASP..111...63F} extinction law to calculate both Galactic and host galaxy extinction. The Galactic extinction parameters are obtained from the IRSA Dust Extinction Service \footnote{\url{https://irsa.ipac.caltech.edu/applications/DUST/}}, with $R_V$ = 3.1 and $E(B-V)_{\rm host}$ = 0.053. For the host galaxy extinction, the $R_V$ of the SMC is assumed \citep{2006ApJ...641..993K,2010MNRAS.401.2773S}, while $E(B-V)$ is treated as a free parameter in the parameter inference.

\begin{figure*}
    \centering
    \includegraphics[width=0.85\textwidth]{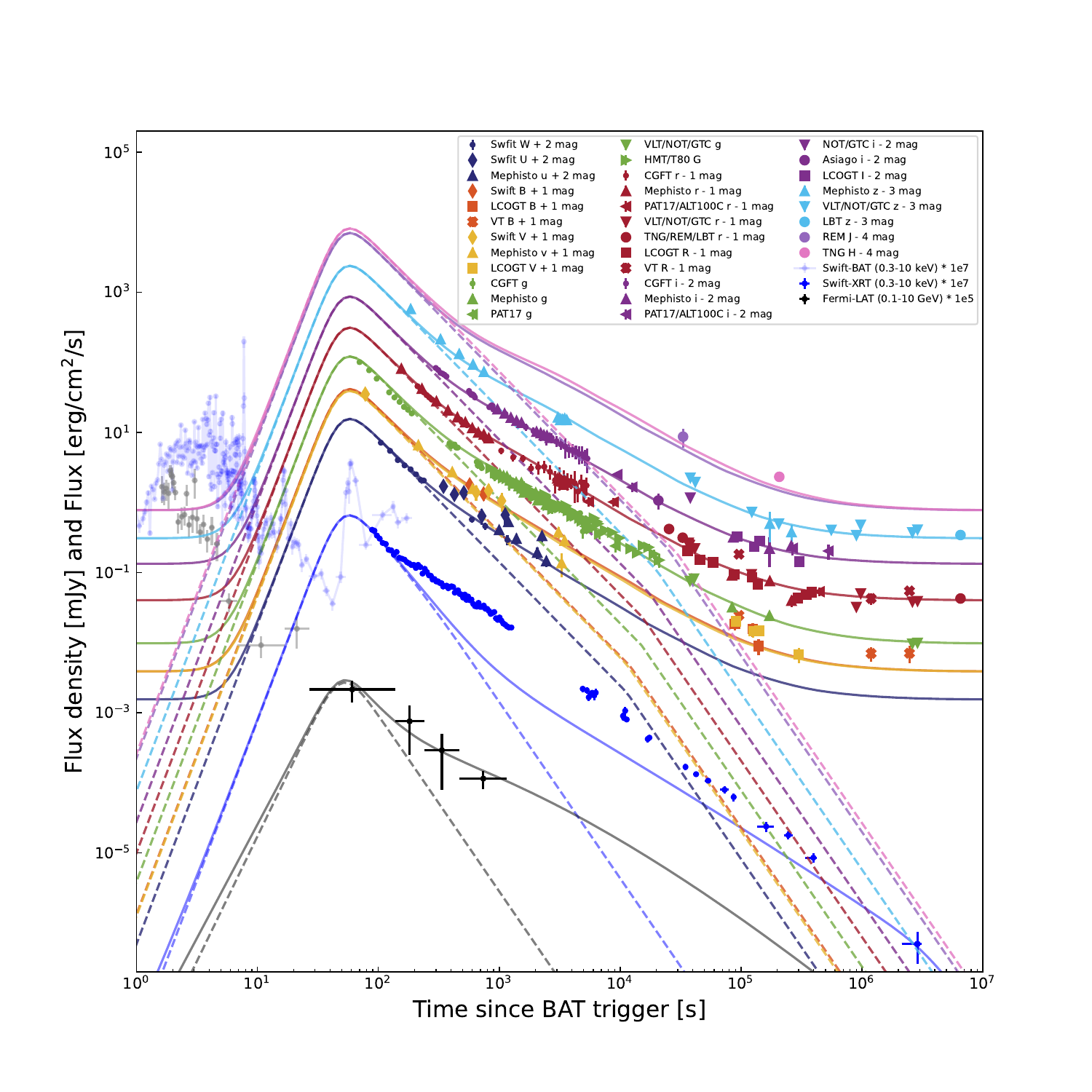}
    \caption{The multiband observations of GRB 240825A and the predicted values from the afterglow model with optimal parameters. The solid line represents the model line ($F_{t, \nu}$),
    which is the sum of the RS, FS and host galaxy emissions, and the red dashed line represents the RS component. Considering that the early data are related to the prompt phase, the black and blue data points with high transparency are excluded from the fitting. The observational upper limits were not taken into account in the fitting. The observed data and model line in the figure both account for extinction correction.}
    \label{fig:fs_rs}
\end{figure*}

\begin{figure*}[htp!]
    \centering
    \includegraphics[width=0.45\textwidth]{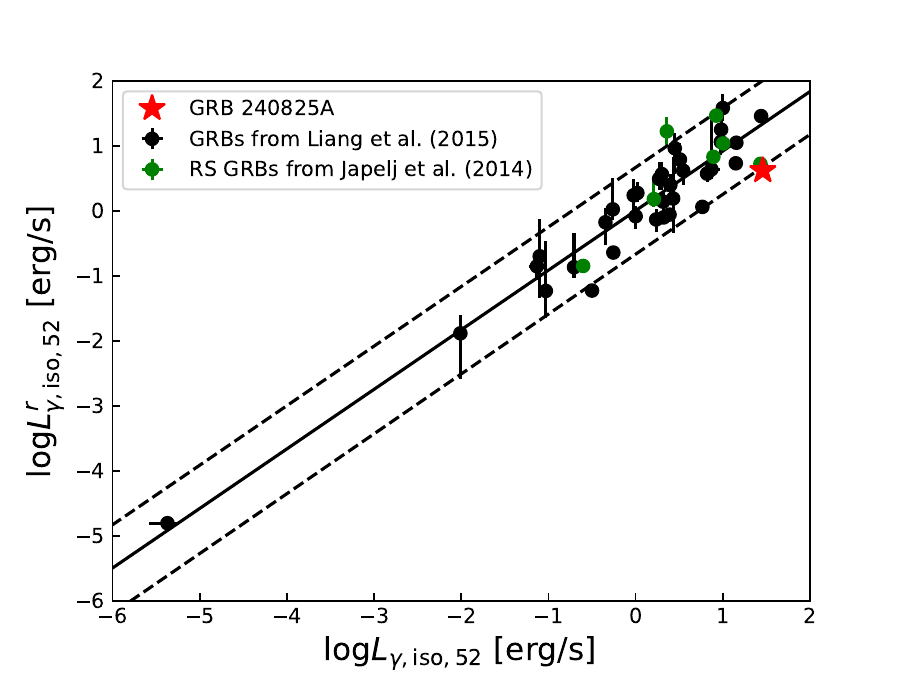}
    \includegraphics[width=0.45\textwidth]{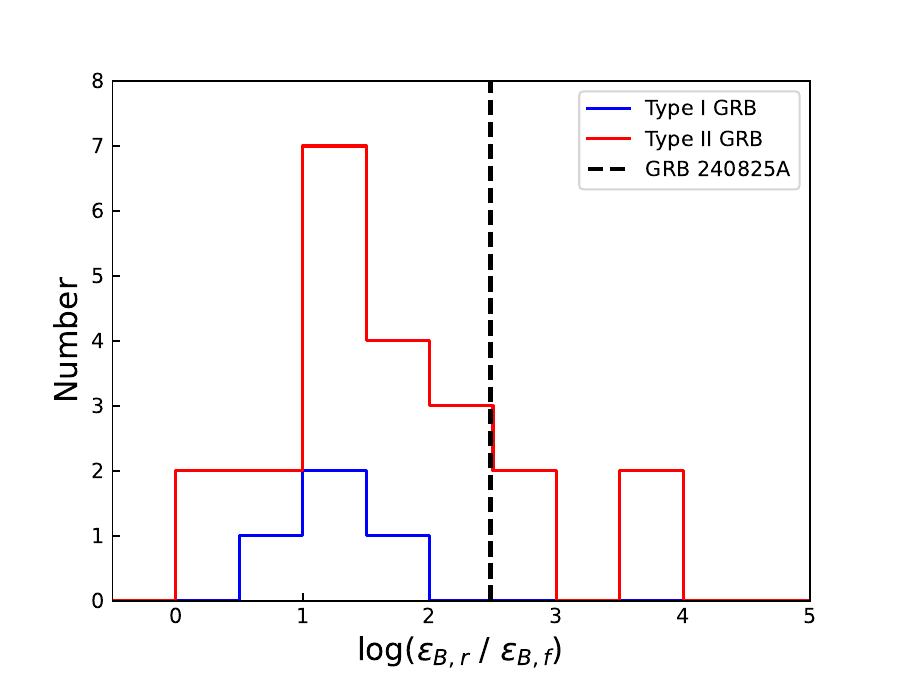}
    \caption{Comparison plots with statistical studies. {\it Left:} The compact three-parameter relation and the black dots are from \cite{2015ApJ...813..116L}, the green dots come from \cite{2014ApJ...785...84J}, and the red star represents GRB 240825A. The solid and dashed lines mark the relation and its 2$\sigma$ dispersion respectively. {\it Right:} The histogram comes from the statistical work of \cite{2024ApJ...969..146W} on RS GRBs. The black dashed line represents GRB 240825A.}
    \label{fig:ll}
\end{figure*}

\subsection{Parameter inference}
We use {\tt PyMultiNest} \citep{2014A&A...564A.125B} as a sampler to perform Bayesian inference on the parameters of the afterglow model (Eq. \ref{eq:fs_rs}), with the number of live points set to 500. The prior ranges for these parameters are listed in Table \ref{tab_param}, where the prior ranges of $E_{\rm k,iso}$ and $\Gamma_{0}$ take into account the constraints from the prompt emission analysis results. We consider a $\gamma$-ray radiation efficiency $\eta_{\gamma} >$ 0.1$\%$ (the minimum value reported in the statistical study by \cite{2007ApJ...655..989Z}), which corresponds to an upper bound of 56.23 for the prior of $\log_{10} E_{\mathrm{k,iso}}$.
\wyAdd{For prompt emission with a distinct high-energy cutoff, if it is caused by $\gamma\gamma$ absorption, the Lorentz factor can be estimated as $\Gamma_{\rm cut} \sim (1+z){E_{\rm cutoff}}/{m_e c^2}$ \citep{2010ApJ...709..525L,2015ApJ...806..194T}. Considering that the dissipation mechanism during the prompt phase may affect the Lorentz factor, this estimate is treated as an upper limit, corresponding to an upper boundary of 2.39 for the prior on $\log_{10} \Gamma_0$ in the afterglow fitting. However, if the high-energy cutoff originates intrinsically, a different approach would be required to constrain the Lorentz factor. Therefore, our estimate here is quite rough and should be regarded merely as an attempt to apply one possible method for constraining the Lorentz factor in the absence of a detected afterglow onset.}
Furthermore, we defined a log-likelihood term for the $i$-th data point, representing an observation at time $t_i$ in a band with a central frequency $\nu_i$, as:
\begin{equation}
    \ln \mathcal{L}_i=-\frac{1}{2} \frac{(F_{t, \nu} - F_{t, \nu,\rm obs})^2}{\sigma_i^2+f_{sys}^2F_{t, \nu,\rm obs}^2} - \ln [2 \pi (\sigma_i^2+f_{sys}^2F_{t, \nu,\rm obs}^2)].
\end{equation}
Considering that the inference involves observational data from various instruments, we introduce a free parameter $f_{sys}$ to characterize the systematic error and normalize the data in each band by their maximum value (i.e. $F_{t, \nu}$ and $F_{t, \nu,\rm obs}$ are normalized values). The total log-likelihood function is expressed as $\ln \mathcal{L} = \sum \ln \mathcal{L}_i$. The posterior parameters obtained from the Bayesian inference, along with their 1$\sigma$ confidence intervals, are listed in Table \ref{tab_param}. Figure \ref{fig:fs_rs_conrner} in the Appendix presents the corner plot of the parameter distributions.

The multiband observational data and the model predictions based on the best-fit parameters are shown in Figure \ref{fig:fs_rs}. 
The FS+RS afterglow model provides a good explanation for the optical data and is roughly consistent with the GeV observations (despite large uncertainties), but it does not explain the X-ray observations well, suggesting that there may be an additional contribution to the X-ray emission.
By combining the afterglow model and prompt emission to constrain the initial Lorentz factor, we can examine the tight parameter correlation proposed by \cite{2015ApJ...813..116L}, i.e. $L^{\rm r}_{\rm iso,52}=-(6.38\pm0.35)+(1.34\pm0.14)\times\log(E_{\rm p,z}/{\rm keV})+(1.32\pm0.19)\times\log\Gamma_0$. As illustrated in the left side of Figure \ref{fig:ll}, GRB 240825A generally aligns with this correlation and, similar to other GRBs with RS emission (hereafter referred to as RS GRBs) \citep{2014ApJ...785...84J}, exhibits higher luminosity compared to other typical Type II GRBs. Furthermore, based on the microphysics parameters of magnetic energy in the FS and RS obtained from the posterior parameters, we can estimate the magnetic ratio of RS and FS jet with $\mathcal{R}_B= {\epsilon_{B,r}}/{\epsilon_{B,f}}=302$\footnote{$\mathcal{R}_B \equiv {B_r}/{B_f}=({\epsilon_{B,r}}/{\epsilon_{B,f}})^{1/2}$ is defined in \cite{2005ApJ...628..315Z}. To compare with some statistical studies \citep{2014ApJ...785...84J,2020ApJ...895...94Y,2024ApJ...969..146W}, we adopt $\mathcal{R}_B= {\epsilon_{B,r}}/{\epsilon_{B,f}}$ here, which requires taking the square root for conversion to the magnetic field ratio.}. As shown on the right side of Figure \ref{fig:ll}, GRB 240825A exhibits a RS-to-FS magnetic ratio that is in the high end of the distribution derived by \citep{2024ApJ...969..146W}, and the $E(B-V) = 0.37^{+0.02}_{-0.02}$ of the host galaxy we finally obtained as free parameters matches well the constraint obtained by \cite{2025ApJ...979...38C}.

\section{Summary and Discussion}\label{sec:sd}
In this work, we performed a detailed analysis of the prompt emission spectrum using data from Swift-BAT and Fermi-GBM/LAT and modeled the multiwavelength (optical/NIR, X-ray, $\gamma$-ray) afterglow light curves using the FS+RS model. By combining the observational data with our analysis results, we summarize the characteristics of this event as follows:
\begin{itemize}
\item{The prompt emission spectrum of GRB 240825A is best fit by a model combining a Band function and a power-law component, with a high-energy cutoff of $\sim$ 76 MeV. The peak energy ($E_{\rm p}$) of the Band function is approximately 500 keV, and the $k$-corrected isotropic energy $E_{\gamma,\rm iso}$ is estimated to be $\sim$ 1.7 $\times$ $10^{53}$ erg. These values make GRB 240825A consistent with the Amati relation. } 

\item{
Considering the high-energy cutoff of the prompt emission due to $\gamma\gamma$ absorption and the radiation efficiency ($\eta_{\gamma} >$ 0.1$\%$), constraints yield an initial Lorentz factor $\Gamma_0 <$ 245 and an isotropic kinetic energy $E_{\text{k,iso}} <$ 1.7 $\times$ $10^{56}$ erg. These two constraints from prompt emission were used to update the prior ranges of parameters during Bayesian inference in the afterglow modeling.}
\item{Based on the current afterglow fitting results with the FS+RS model, the observations can be effectively explained in both the optical and GeV bands. However, the current model fails to explain the X-ray observations. The initial Lorentz factor derived from the afterglow modelling (\wyAdd{$\Gamma_{0} \sim 234$}) is consistent with the constraint derived from the high-energy cutoff of the prompt emission (\wyAdd{$\Gamma_{\rm cut} \sim 245$}). This is expected, as in the absence of a detected optical onset in the afterglow, the constraint on the Lorentz factor largely depends on the prior set based on the high-energy cutoff observed in the prompt emission. We obtain a RS-to-FS magnetic ratio ($\mathcal{R}_B$) of 302, and a radiation efficiency in the $\gamma$-rays ($\eta_{\gamma}$) of 3.1\%. The large value of $\mathcal{R}_B \gg 1$ is consistent with the scenario in which the RS makes a significant contribution to the observed emission.}
\item{Comparing GRB 240825A with previous empirical relations and statistical studies, it aligns with the Type II GRB region on the $E_{\rm p,z}$\textendash$E_{\gamma,\rm iso}$ relation. Its initial Lorentz factor, energy peak, and isotropic luminosity satisfy the tight three-parameter correlation proposed by \cite{2015ApJ...813..116L}. Moreover, it exhibits a higher $\mathcal{R}_B$ compared to other long GRBs \citep{2024ApJ...969..146W}.}
\end{itemize}

Based on the aforementioned characteristics of GRB~240825A, we have conducted discussions on the following points. Firstly, GRB 240825A is an event with comprehensive observational coverage across time and wavelength, spanning from the prompt emission to the afterglow phase. During the afterglow phase, the emission is initially dominated by RS components, and transitions to being FS 
dominated, and eventually reflects contributions from the host galaxy. Secondly, such a ``flattening case" in the combination of FS+RS typically requires $\mathcal{R}_B$ $\gg$ 1 \citep{2003ApJ...595..950Z}, which is consistent with the result obtained from our model fitting. This suggests that the RS is more magnetized than the FS, but is not Poynting-flux-dominated (otherwise the RS is suppressed). Secondly, although this burst appears to be a typical Type II GRB based on empirical correlations, its comprehensive observational data make it an excellent probe for studying the physics of GRB jets. For example, the initial Lorentz factor of the afterglow is often correlated with the onset time. Although SVOM/C-GFT conducted rapid follow-up observations of GRB~240825A, the optical rise phase is still missing. Therefore, the constraints on the Lorentz factor derived from the prompt emission are crucial, highlighting the importance of comprehensive observations from the prompt emission to the afterglow phase. 
\wyAdd{Thirdly, despite the overall consistency between the model and the observed multiwavelength afterglow data, the X-ray light curve is not well reproduced in our current framework. This discrepancy may point to additional physical components or emission mechanisms not accounted for in the standard external shock model. Several scenarios have been proposed in the literature to explain the shaping of X-ray afterglow emission in GRBs, such as energy injection, evolving shock parameters, etc.} \citep{2006ApJ...642..354Z,2006MNRAS.369..197F}. However, most of these scenarios generally predict correlated features in other bands, particularly in the optical or high-energy regimes. The fact that our model provides a good fit to the optical, while significantly underpredicting the X-ray flux, suggests that the X-ray excess may arise from an additional component that selectively contributes to the X-ray band. In previous work, such as GRB 130427A, X-rays may come from the contributions of flash-RS and RS \cite{2014Sci...343...38V}. In addition, based on GRB 221009A, people realized that there may be an extremely narrow component in the jet structure \cite{2024JHEAp..41...42Z,2024ApJ...962..115R}, i.e., the two-component jet model. Hence, this remains an open question, and future observations and modeling efforts will be necessary to fully understand the origin of such X-ray behavior.

\begin{acknowledgements}
The Space-based multiband Variable Objects Monitor (SVOM) is a joint Chinese-French mission led by the Chinese National Space Administration (CNSA), the French Space Agency (CNES), and the Chinese Academy of Sciences (CAS). We gratefully acknowledge the unwavering support of NSSC, IAMCAS, XIOPM, NAOC, IHEP, CNES, CEA, and CNRS.

This work is supported by the National Key R\&D Program of China (grant No.~2024YFA1611600), the \textit{SVOM} project (a mission under the Strategic Priority Program on Space Science of the Chinese Academy of Sciences), the Strategic Priority Research Program of the Chinese Academy of Sciences (NFSC, grant No.~XDB0550401), and the National Natural Science Foundation of China (NSFC, grant No.~12494573).
This research was partly supported by Natural Science Foundation of Xinjiang Uygur Autonomous Region (grant No.~2024D01D32), Tianshan Talent Training Program (grant No.~2023TSYCLJ0053), and Tianshan Innovation Team Program (grant No.~2024D14015). 

This work is also supported by the following funding programs:
Y.W. is supported by the Jiangsu Funding Program for Excellent Postdoctoral Talent (grant No.~2024ZB110), the Postdoctoral Fellowship Program (grant No.~GZC20241916), and the General Fund (grant No.~2024M763531) of the China Postdoctoral Science Foundation.  
G.P.L. is supported by a Royal Society Dorothy Hodgkin Fellowship (grant Nos. DHF-R1-221175 and DHF-ERE-221005). 
A.S. acknowledges support by a postdoctoral fellowship from the CNES.
X.H.H. is supported by the National Key R\&D Program of China (grant No.~2024YFA1611702) and the Strategic Priority Research Program of the Chinese Academy of Sciences (grant No.~XDB0550101).  
S.D.V., E.L.F., B.S. acknowledge the support of the French Agence Nationale de la Recherche (ANR), under grant ANR-23-CE31-0011 (project PEGaSUS). 
R.B., S.C., P.D.A., M.F., A.M., A.Ro., A.Re. acknowledge financial support from the GRAWITA Large 
Program grant (PI P. D’Avanzo) and the PRIN-INAF 2022 "Shedding light on the nature of gap transients: from the observations to the models". 
R.B., S.C., P.D.A., M.F., A.M. acknowledge financial support from the Italian Space Agency, contract ASI/INAF No.~I/004/11/6.
A.Rossi acknowledges support from the INAF project Premiale Supporto Arizona \& Italia. 
E.W.L. is supported by the National Natural Science Foundation of China (NSFC, grant No.~12133003).
X.G.W. is supported by the National Natural Science Foundation of China (NSFC, grant No.~12373042) and the Bagui Scholars Program (No.~GXR-6BG2424001).
D.B.M. is funded by the European Union (ERC, HEAVYMETAL, 101071865, Views and opinions expressed are, however, those of the authors only and do not necessarily reflect those of the European Union or the European Research Council. Neither the European Union nor the granting authority can be held responsible for them). 
The Cosmic Dawn Center (DAWN) is funded by the Danish National Research Foundation under grant No.~DNRF140.
Z.P.J. is supported by the National Natural Science Foundation of China (NSFC, grant Nos.~12225305 and 12321003).
D.M.W.is supported by the National Natural Science Foundation of China (NSFC, grant No. 12473049.
J.R. is supported by the General Fund (grant No.~2024M763530) of the China Postdoctoral Science Foundation.

We acknowledge the following observational support for this work: partly collected at the Schmidt Telescope of the INAF – Osservatorio Astronomico di Padova (Asiago, Italy) and with the Italian Telescopio Nazionale Galileo (TNG) operated on La Palma by the Fundaci\'{o}n  Galileo Galilei of the INAF at the Observatorio del Roque de los Muchachos of the Instituto de Astrof\'isica  de Canarias; 
partly based on data obtained with the instrument OSIRIS, built by a Consortium led by the Instituto de Astrof\'isica de Canarias in collaboration with the Instituto de Astronom\'ia of the Universidad Aut\'onoma de México. OSIRIS was funded by GRANTECAN and the National Plan of Astronomy and Astrophysics of the Spanish Government (program: GTCMULTIPLE2F-24A, PI: Ag\"u\'i Fern\'andez);
partly collected at the European Organisation for Astronomical Research in the Southern Hemisphere under ESO programme 110.24CF (PI Tanvir, Vergani, Malesani), and with the Nordic Optical Telescope (program 69-023; PI: Malesani), owned in collaboration by the University of Turku and Aarhus University, and operated jointly by Aarhus University, the University of Turku,  and the University of Oslo (representing Denmark, Finland and Norway), the University of Iceland and Stockholm University; partly based on observations made at Observatoire de Haute Provence (CNRS), France, with MISTRAL. We also acknowledge the support from the staff at Nanshan Station of Xinjiang Astronomical Observatory (XAO) for their assistance with the Photometric Auxiliary Telescope (PAT-17). 

This research has made use of the MISTRAL database, operated at CeSAM (LAM), Marseille, France.
This work made use of data supplied by the UK Swift Science Data Centre at the University of Leicester. We acknowledge the use of the Swift and Fermi archive's public data.
\end{acknowledgements}

\appendix
\section{Afterglow observation data and spectral fitting results}
\begin{figure}[htbp]
    \centering
    \includegraphics[width=0.95\textwidth]{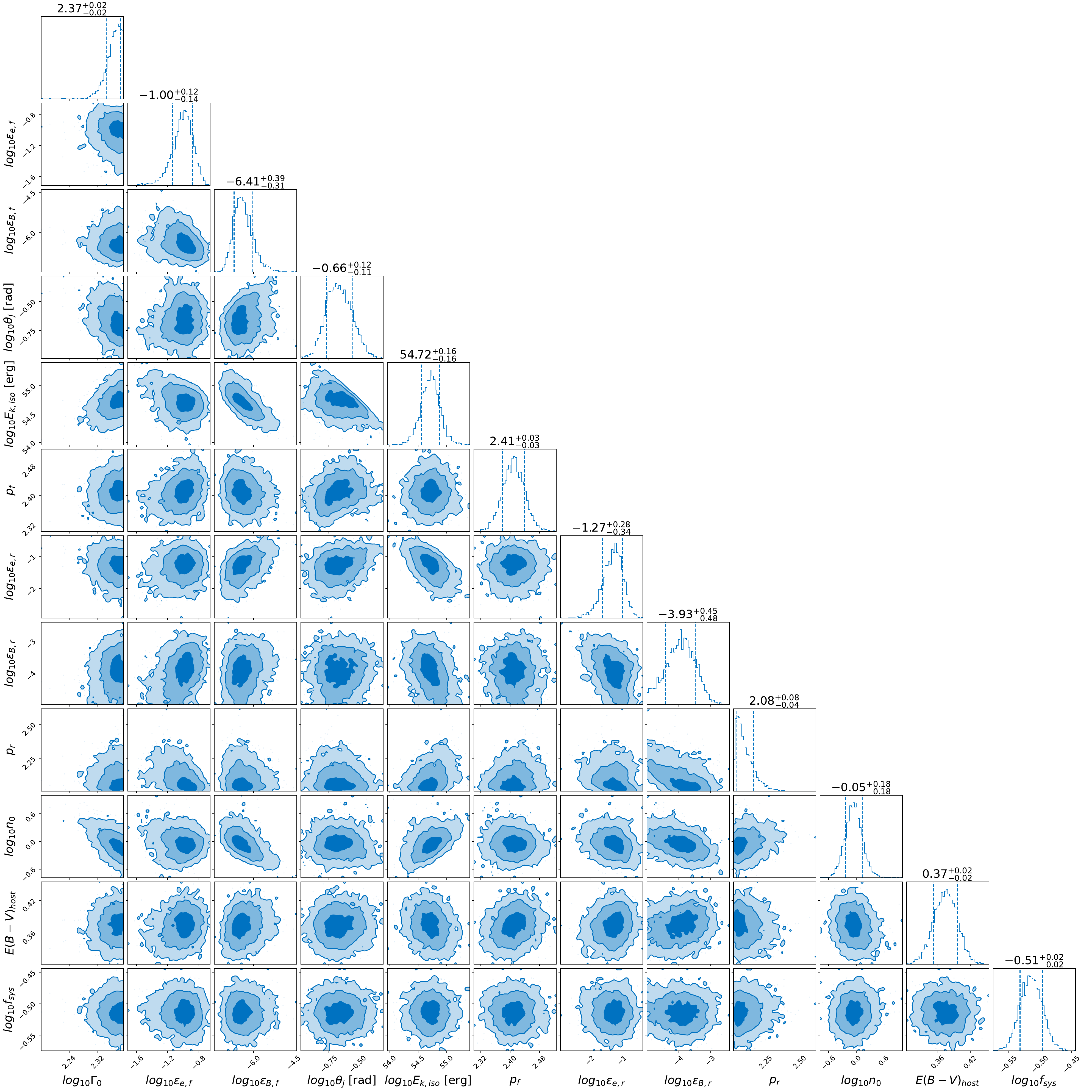}
    \caption{The corner plot of the posterior parameters of the afterglow model.}
    \label{fig:fs_rs_conrner}
\end{figure}

In this appendix, we provide detailed information on the observation data and fitting results of afterglow. The observation data from various telescopes are listed as follows: SVOM C-GFT (Table \ref{tab_cgft}), SVOM VT (Table \ref{tab_vt}), Swift UVOT (Table \ref{tab_uvot}), HMT/T80 (Table \ref{tab_xm}), PAT17/ALT100C (Table \ref{tab_xm2}), LCOGT (Table \ref{tab_LCOGT}), VLT/NOT/GTC (Table \ref{tab_NOTVLT}), and Asiago/REM/TNG/LBT/MISTRAL (Table \ref{tab_REM}). Table \ref{tab_param} presents the priors and posteriors of the afterglow model parameters. Figure \ref{fig:fs_rs_conrner} shows the corner plot of the posterior distributions for the afterglow fitting, and the corresponding parameter values, along with their prior ranges, are listed in Table \ref{tab_param}.

\begin{longtable}[c]{rcccc} 
\caption{SVOM C-GFT observations} \label{tab_cgft} \\
  \hline\noalign{\smallskip}
Mid time (s) & Exposure time (s)  & Mag (AB)  & Mag$_{\rm err}$ & Filter \\
  \hline\noalign{\smallskip}
\endfirsthead 
\multicolumn{5}{c}{SVOM C-GFT observations (Continued)} \\ \hline
Mid Time (s) & Exposure Time (s)  &  Mag (AB)   & Mag$_{\rm err}$ & Filter\\
 \hline\noalign{\smallskip}
\endhead 
\hline
\endfoot 
\endlastfoot 
71   &  10  &13.74  & 0.02 &  g   \\ 
84   &  10  &14.03  & 0.02 &  g   \\ 
97   &  10  &14.33  & 0.02 &  g   \\ 
123  &  10  &14.82  & 0.02 &  g   \\ 
136  &  10  &15.00  & 0.02 &  g   \\ 
150  &  10  &15.16  & 0.02 &  g   \\ 
163  &  10  &15.34  & 0.03 &  g   \\ 
176  &  10  &15.44  & 0.03 &  g   \\ 
189  &  10  &15.57  & 0.03 &  g   \\ 
388  &  10  &16.64  & 0.05 &  g   \\ 
401  &  10  &16.74  & 0.06 &  g   \\ 
414  &  10  &16.72  & 0.06 &  g   \\ 
427  &  10  &16.68  & 0.06 &  g   \\ 
441  &  10  &16.79  & 0.06 &  g   \\ 
454  &  10  &16.79  & 0.06 &  g   \\ 
652  &  10  &17.30  & 0.09 &  g   \\ 
665  &  10  &17.43  & 0.10 &  g   \\ 
678  &  10  &17.35  & 0.10 &  g   \\ 
692  &  10  &17.37  & 0.10 &  g   \\ 
705  &  10  &17.44  & 0.10 &  g   \\ 
718  &  10  &17.52  & 0.11 &  g   \\ 
917  &  10  &17.84  & 0.15 &  g   \\ 
930  &  10  &17.79  & 0.14 &  g   \\ 
944  &  10  &17.65  & 0.12 &  g   \\ 
957  &  10  &17.72  & 0.13 &  g   \\ 
970  &  10  &17.74  & 0.13 &  g   \\ 
984  &  10  &17.72  & 0.13 &  g   \\ 
1216 &  $^{*}6\times10$  &18.02  & 0.26 &   g   \\
1594 &  $^{*}5\times10$  &18.22  & 0.26 &   g   \\  
1753 &  $^{\blacktriangle}5\times10$  &18.50  & 0.12 &  g   \\  
2011 &  $^{\blacktriangle}6\times10$  &18.44  & 0.10 &  g   \\  
2276 &  $^{\blacktriangle}6\times10$  &18.59  & 0.10 &  g   \\  
2540 &  $^{\blacktriangle}6\times10$  &18.54  & 0.09 &  g   \\  
2806 &  $^{\blacktriangle}6\times10$  &18.76  & 0.11 &  g   \\  
3071 &  $^{\blacktriangle}6\times10$  &18.86  & 0.11 &  g   \\  
3336 &  $^{\blacktriangle}6\times10$  &19.02  & 0.11 &  g   \\  
3601 &  $^{\blacktriangle}6\times10$  &19.08  & 0.12 &  g   \\  
3866 &  $^{\blacktriangle}6\times10$  &19.09  & 0.14 &  g   \\  
4131 &  $^{\blacktriangle}6\times10$  &19.18  & 0.15 &  g   \\  
4396 &  $^{\blacktriangle}6\times10$  &19.31  & 0.14 &  g   \\  
4661 &  $^{\blacktriangle}6\times10$  &19.13  & 0.12 &  g   \\  
4925 &  $^{\blacktriangle}6\times10$  &19.78  & 0.23 &  g   \\  
5190 &  $^{\blacktriangle}6\times10$  &19.37  & 0.16 &  g   \\  
211  &  10  &15.11  & 0.02 &  r  \\  
224  &  10  &15.23  & 0.02 &  r  \\  
238  &  10  &15.33  & 0.02 &  r  \\  
251  &  10  &15.43  & 0.02 &  r  \\  
264  &  10  &15.51  & 0.02 &  r  \\  
277  &  10  &15.59  & 0.02 &  r  \\  
476  &  10  &16.30  & 0.04 &  r  \\  
489  &  10  &16.39  & 0.04 &  r  \\  
502  &  10  &16.44  & 0.04 &  r  \\  
515  &  10  &16.44  & 0.04 &  r  \\  
529  &  10  &16.48  & 0.04 &  r  \\  
542  &  10  &16.50  & 0.05 &  r  \\  
740  &  10  &16.84  & 0.06 &  r  \\  
754  &  10  &16.98  & 0.06 &  r  \\  
767  &  10  &17.03  & 0.07 &  r  \\  
780  &  10  &16.99  & 0.07 &  r  \\  
793  &  10  &17.00  & 0.07 &  r  \\  
806  &  10  &16.97  & 0.07 &  r  \\  
1039 &  $^{*}6\times10$  &17.41  & 0.12 &  r  \\ 
1304 &  $^{*}6\times10$  &17.65  & 0.14 &  r  \\ 
1570 &  $^{*}6\times10$  &17.69  & 0.13 &  r  \\ 
1835 &  $^{*}6\times10$  &18.04  & 0.19 &  r  \\ 
2100 &  $^{*}6\times10$  &18.00  & 0.23 &  r  \\ 
2364 &  $^{*}6\times10$  &17.98  & 0.22 &  r  \\ 
2629 &  $^{*}6\times10$  &18.16  & 0.25 &  r  \\ 
2894 &  $^{*}6\times10$  &18.46  & 0.39 &  r  \\ 
3159 &  $^{*}6\times10$  &18.34  & 0.40 &  r  \\ 
3424 &  $^{*}6\times10$  &18.40  & 0.32 &  r  \\ 
3689 &  $^{*}6\times10$  &18.48  & 0.30 &  r  \\ 
3954 &  $^{*}6\times10$  &18.50  & 0.26 &  r  \\ 
4219 &  $^{*}6\times10$  &18.57  & 0.52 &  r  \\ 
4484 &  $^{*}6\times10$  &18.87  & 0.33 &  r  \\ 
4749 &  $^{*}6\times10$  &18.53  & 0.29 &  r  \\ 
5013 &  $^{*}6\times10$  &18.75  & 0.36 &  r  \\ 
5278 &  $^{*}6\times10$  &18.65  & 0.31 &  r  \\ 
299  &  10  &15.14  & 0.02 &  i  \\   
313  &  10  &15.21  & 0.02 &  i  \\   
326  &  10  &15.31  & 0.02 &  i  \\   
339  &  10  &15.34  & 0.02 &  i  \\   
352  &  10  &15.38  & 0.02 &  i  \\   
366  &  10  &15.44  & 0.02 &  i  \\   
563  &  10  &15.96  & 0.03 &  i  \\   
577  &  10  &16.05  & 0.04 &  i  \\   
590  &  10  &16.10  & 0.04 &  i  \\   
603  &  10  &16.08  & 0.04 &  i  \\   
617  &  10  &16.12  & 0.04 &  i  \\   
630  &  10  &16.19  & 0.04 &  i  \\   
829  &  10  &16.54  & 0.05 &  i  \\   
842  &  10  &16.53  & 0.06 &  i  \\   
855  &  10  &16.49  & 0.05 &  i  \\   
868  &  10  &16.47  & 0.05 &  i  \\   
882  &  10  &16.49  & 0.05 &  i  \\   
895  &  10  &16.62  & 0.06 &  i  \\  
1127  &  $^{*}6\times10$  &16.78  & 0.08 &  i  \\  
1379  &  $^{*}4\times10$  &17.04  & 0.11 &  i  \\  
1658  &  $^{*}6\times10$  &17.25  & 0.12 &  i  \\  
1923  &  $^{*}6\times10$  &17.39  & 0.14 &  i  \\  
2187  &  $^{*}6\times10$  &17.51  & 0.21 &  i  \\  
2452  &  $^{*}6\times10$  &17.66  & 0.22 &  i  \\  
2717  &  $^{*}6\times10$  &17.63  & 0.19 &  i  \\  
2982  &  $^{*}6\times10$  &17.75  & 0.19 &  i  \\  
3248  &  $^{*}6\times10$  &17.86  & 0.19 &  i  \\  
3513  &  $^{*}6\times10$  &17.94  & 0.35 &  i  \\  
3778  &  $^{*}6\times10$  &18.06  & 0.26 &  i  \\  
4042  &  $^{*}6\times10$  &18.06  & 0.24 &  i  \\  
4308  &  $^{*}6\times10$  &18.12  & 0.24 &  i  \\  
4573  &  $^{*}6\times10$  &18.21  & 0.31 &  i  \\  
4837  &  $^{*}6\times10$  &18.21  & 0.30 &  i  \\  
5101  &  $^{*}6\times10$  &18.32  & 0.28 &  i  \\  
5367  &  $^{*}5\times10$  &18.37  & 0.41 &  i  \\ 
  \hline\noalign{\smallskip}
 \end{longtable}
  \vspace{-10pt}
  {\footnotesize
  Notes: Data binning or image stacking was employed to mitigate the effects of outliers or enhance the signal-to-noise ratio for the data of faint  afterglow at late phase. ``${*}$" indicates binned data. ``${\blacktriangle}$" indicates stacked image data.  }

\clearpage
\begin{table}
\begin{center}
\caption[]{SVOM VT observations}\label{tab_vt}
 \begin{tabular}{rcccc}
  \hline\noalign{\smallskip}
Mid Time (s) &  Exposure Time (s)      & Mag (AB) & Mag$_{\rm err}$ & Filter                   \\
  \hline\noalign{\smallskip}
96,232     &  $102\times20$ & 22.09	& 0.10 & VT\_B \\
96,262     &  $ 95\times20$ & 20.90	& 0.05 & VT\_R \\
1,192,978   &  $104\times20$ & 23.45	& 0.25 & VT\_B \\
1,192,968   &  $143\times20$ & 22.48	& 0.20 & VT\_R \\
2,472,972   &  $199\times20$ & 23.44	& 0.30 & VT\_B \\ 
2,472,973   &  $196\times20$ & 22.21	& 0.10 & VT\_R \\
  \noalign{\smallskip}\hline
\end{tabular}
\end{center}
\end{table}

\begin{table}
\begin{center}
\caption[]{Swift UVOT observations}\label{tab_uvot}
\begin{tabular}{rcccc}
  \hline\noalign{\smallskip}
Mid Time (s) &  Exposure Time (s)      & Mag (AB) & Mag$_{\rm err}$ & Filter   \\
  \hline\noalign{\smallskip}
78 &	9.8 &	13.61 &	0.07 &	V       \\
105 &	19.7 &	15.64 &	0.04 &	WHITE   \\
125 &	19.7 &	15.93 &	0.04 &	WHITE   \\
145 &	19.7 &	16.26 &	0.04 &	WHITE   \\
165 &	19.7 &	16.45 &	0.04 &	WHITE   \\
185 &	19.7 &	16.59 &	0.05 &	WHITE   \\
205 &	19.7 &	16.77 &	0.05 &	WHITE   \\
225 &	19.7 &	16.97 &	0.05 &	WHITE   \\
238 &	6.9 &	16.99 &	0.08 &	WHITE   \\
346 &	78.7 &	17.72 &	0.08 &	U       \\
426 &	78.7 &	18.01 &	0.09 &	U       \\
510 &	87.6 &	17.95 &	0.08 &	U       \\
570 &	18.7 &	17.45 &	0.18 &	B       \\
594 &	18.7 &	18.38 &	0.10 &	WHITE   \\
644 &	18.7 &	17.12 &	0.26 &	V       \\
718 &	18.7 &	18.78 &	0.28 &	U       \\
743 &	17.7 &	17.77 &	0.22 &	B       \\
767 &	18.7 &	18.62 &	0.12 &	WHITE   \\
817 &	17.7 &	17.06 &	0.26 &	V       \\
1048 &	18.7 &	17.42 &	0.32 &	V       \\
1121 &	18.7 &	18.74 &	0.31 &	U       \\   
1170 &	18.7 &	19.11 &	0.18 &	WHITE   \\
\noalign{\smallskip}\hline \\
\end{tabular}
\end{center}
\tablecomments{0.86\textwidth}{All observation data with WHITE filter and the last two points with U filter are affected by the SSS effect.}
\end{table}

\begin{longtable}[c]{rccccc} %
\caption{HMT/T80 observations}\label{tab_xm}  \\
  \hline\noalign{\smallskip}
Mid Time (s) &  Exposure Time (s)      & Mag & Mag$_{\rm err}$ & Filter & Telescope \\  
  \hline\noalign{\smallskip}
\endfirsthead 
\multicolumn{6}{c}{HMT/T80 observations (Continued)} \\ \hline
Mid Time (s) &  Exposure Time (s)      & Mag & Mag$_{\rm err}$ & Filter & Telescope \\  
 \hline\noalign{\smallskip}
\endhead %
\hline
\endfoot 
\endlastfoot 
849 & $60$ & 17.00  & 0.03  & clear & HMT \\ 
925 & $60$ & 17.16  & 0.04  & clear & HMT \\ 
1001 & $60$ & 17.26  & 0.04  & clear & HMT \\ 
1077 & $60$ & 17.35  & 0.04  & clear & HMT \\ 
1154 & $60$ & 17.37  & 0.05  & clear & HMT \\ 
1230 & $60$ & 17.42  & 0.05  & clear & HMT \\ 
1306 & $60$ & 17.48  & 0.05  & clear & HMT \\ 
1383 & $60$ & 17.51  & 0.06  & clear & HMT \\ 
1459 & $60$ & 17.60  & 0.06  & clear & HMT \\ 
1536 & $60$ & 17.73  & 0.08  & clear & HMT \\ 
1613 & $60$ & 17.71  & 0.08  & clear & HMT \\ 
1689 & $60$ & 17.65  & 0.08  & clear & HMT \\ 
1766 & $60$ & 17.80  & 0.08  & clear & HMT \\ 
1842 & $60$ & 17.80  & 0.08  & clear & HMT \\ 
1919 & $60$ & 17.97  & 0.09  & clear & HMT \\ 
1995 & $60$ & 18.00  & 0.09  & clear & HMT \\ 
2071 & $60$ & 17.91  & 0.08  & clear & HMT \\ 
2147 & $60$ & 17.95  & 0.08  & clear & HMT \\ 
2224 & $60$ & 18.00  & 0.09  & clear & HMT \\ 
2301 & $60$ & 18.03  & 0.09  & clear & HMT \\ 
2377 & $60$ & 17.94  & 0.08  & clear & HMT \\ 
2454 & $60$ & 18.12  & 0.09  & clear & HMT \\ 
2530 & $60$ & 18.33  & 0.12  & clear & HMT \\ 
2607 & $60$ & 18.26  & 0.12  & clear & HMT \\ 
2683 & $60$ & 18.28  & 0.14  & clear & HMT \\ 
2759 & $60$ & 18.46  & 0.15  & clear & HMT \\ 
2836 & $60$ & 18.40  & 0.16  & clear & HMT \\ 
2912 & $60$ & 18.29  & 0.13  & clear & HMT \\ 
2989 & $60$ & 18.30  & 0.12  & clear & HMT \\ 
3155 & $120$ & 18.41  & 0.11  & clear & HMT \\ 
3292 & $120$ & 18.33  & 0.10  & clear & HMT \\ 
3428 & $120$ & 18.41  & 0.11  & clear & HMT \\ 
3565 & $120$ & 18.34  & 0.10  & clear & HMT \\ 
3701 & $120$ & 18.51  & 0.13  & clear & HMT \\ 
3838 & $120$ & 18.49  & 0.14  & clear & HMT \\ 
3974 & $120$ & 18.71  & 0.14  & clear & HMT \\ 
4111 & $120$ & 18.59  & 0.11  & clear & HMT \\ 
4247 & $120$ & 18.70  & 0.12  & clear & HMT \\ 
4384 & $120$ & 18.63  & 0.11  & clear & HMT \\ 
4615 & $200$ & 18.78  & 0.13  & clear & HMT \\ 
4832 & $200$ & 18.88  & 0.14  & clear & HMT \\ 
5048 & $200$ & 18.82  & 0.12  & clear & HMT \\ 
5264 & $200$ & 18.86  & 0.11  & clear & HMT \\ 
5481 & $200$ & 19.06  & 0.14  & clear & HMT \\ 
5776 & $300$ & 19.16  & 0.14  & clear & HMT \\ 
6092 & $300$ & 18.78  & 0.10  & clear & HMT \\ 
6409 & $300$ & 18.90  & 0.11  & clear & HMT \\ 
6725 & $300$ & 19.35  & 0.19  & clear & HMT \\ 
5647.5 & $3\times300$ & 18.88  & 0.03  & clear & T80 \\ 
6585.0 & $3\times300$ & 18.98  & 0.03  & clear & T80 \\ 
7523.5 & $3\times300$ & 19.21  & 0.04  & clear & T80 \\ 
8617.5 & $4\times300$ & 19.29  & 0.04  & clear & T80 \\ 
9868.0 & $4\times300$ & 19.49  & 0.05  & clear & T80 \\ 
12,253.0 & $5\times300$ & 19.87  & 0.08  & clear & T80 \\ 
13,889.5 & $5\times300$ & 20.02  & 0.08  & clear & T80 \\ 
15,507.0 & $5\times300$ & 19.76  & 0.09  & clear & T80 \\ 
17,228.0 & $5\times300$ & 19.82  & 0.13  & clear & T80 \\ 
19,115.5 & $6\times300$ & 20.06  & 0.16  & clear & T80 \\ 
21,063.5 & $6\times300$ & 20.28  & 0.24  & clear & T80 \\ 
\noalign{\smallskip}\hline \\
 \end{longtable}

\begin{longtable}[c]{rccccc} %
\caption{PAT17/ALT100C observations}\label{tab_xm2}  \\
  \hline\noalign{\smallskip}
Mid time (s) &  Exposure time (s)      & Mag (AB) & Mag$_{\rm err}$ & Filter & Telescope \\  
  \hline\noalign{\smallskip}
\endfirsthead 
\multicolumn{6}{c}{PAT17/ALT100C observations (Continued)} \\ \hline
Mid time (s) &  Exposure time (s)      & Mag (AB) & Mag$_{\rm err}$ & Filter & Telescope \\  
 \hline\noalign{\smallskip}
\endhead %
\hline
\endfoot 
\endlastfoot 
5168.6 & $600$ & 19.60  & 0.23  & g & PAT17 \\ 
9147.7 & $600$ & 20.30  & 0.22  & g & PAT17 \\ 
5484.6 & $300$ & 19.23  & 0.17  & r & PAT17 \\ 
8837.4 & $600$ & 19.24  & 0.15  & r & PAT17 \\ 
9471.4 & $4\times600$ & 18.94  & 0.09  & i & PAT17 \\ 
12,592.5 & $3\times600$ & 19.39  & 0.17  & i & PAT17 \\ 
85,109.9 & $10\times300$ & $\textgreater$21.30  & ...  & r & ALT100C \\ 
173,284.8 & $9\times300$ & $\textgreater$21.80  & ...  & r & ALT100C \\ 
263,330.3 & $11\times300$ & $\textgreater$19.70  & ...  & r & ALT100C \\ 
444,598.4 & $10\times600$ & 22.44  & 0.15  & r & ALT100C \\ 
88,171.3 & $10\times300$ & $\textgreater$20.50  & ...  & i & ALT100C \\ 
176,499.0 & $12\times300$ & $\textgreater$21.90  & ...  & i & ALT100C \\ 
267,869.3 & $18\times300$ & 21.57  & 0.24  & i & ALT100C \\ 
528,544.8 & $9\times600$ & 21.67  & 0.27  & i & ALT100C \\ 
91,671.1 & $12\times300$ & $\textgreater$20.10  & ...  & z & ALT100C \\ 
180,172.2 & $12\times300$ & $\textgreater$21.10  & ...  & z & ALT100C \\ 
272,560.9 & $12\times600$ & $\textgreater$21.20  & ...  & z & ALT100C \\ 
438,707.2 & $9\times600$ & $\textgreater$21.00  & ...  & z & ALT100C \\ 
\noalign{\smallskip}\hline \\
 \end{longtable}

\begin{table}
\begin{center}
\caption[]{LCOGT observations}\label{tab_LCOGT}
\begin{tabular}{rcccc}
  \hline\noalign{\smallskip}
Mid time (s) &  Exposure time (s)      & Mag & Mag$_{\rm err}$ & Filter                   \\
  \hline\noalign{\smallskip}
36,113    & 1500  & 20.51 & 0.07 & R \\
45,631    & 1500  & 20.84 & 0.09 & R \\
58,874    & 1500  & 20.93 & 0.07 & R \\
87,676    & 1500  & 21.36 & 0.09 & R \\
123,423   & 1500  & 21.46 & 0.12 & R \\
138,070.5 & 1500  & 21.71 & 0.10 & R \\
295,102.5 & 2400  & 22.20 & 0.16 & R \\
347,409   & 3000  & 22.09 & 0.11 & R \\
384,887.5 & 3000  & 21.99 & 0.13 & R \\
89,314    & 1500  & 22.51 & 0.22 & B \\
125,063   & 1500  & 22.70 & 0.24 & B \\
139,709.5 & 1500  & 23.30 & 0.28 & B \\
90,952.5  & 1500  & 21.70 & 0.16 & V \\
126,703   & 1500  & 22.05 & 0.20 & V \\
141,351   & 1500  & 22.04 & 0.16 & V \\
300,342   & 2400  & 22.88 & 0.27 & V \\
92,591    & 1500  & 20.28 & 0.12 & I \\
128,342.5 & 1500  & 20.64 & 0.14 & I \\
142,990   & 1500  & 20.44 & 0.17 & I \\
302,960.5 & 2400  & 21.17 & 0.14 & I \\
\noalign{\smallskip}\hline \\
\end{tabular}
\end{center}
\end{table}

\begin{table}
\begin{center}
\caption[]{VLT/NOT/GTC observations}\label{tab_NOTVLT}
\begin{tabular}{rccccc}
  \hline\noalign{\smallskip}
Mid time (s) &  Exposure time (s)      & Mag (AB) & Mag$_{\rm err}$ & Filter &Telescope                  \\
  \hline\noalign{\smallskip}
37,839.7 & $3\times60$ & 21.55	& 0.05 & g & VLT/XSHOOTER\\
37,605.6	& $3\times60$ & 20.78 & 0.04 & r & VLT/XSHOOTER\\
38,075.6	& $3\times60$ & 19.77 & 0.04 & z & VLT/XSHOOTER\\
40,787.7 & $3\times120$ & 21.47  & 0.07  & g & NOT \\ 
41,488.4 & $2\times120$ & 20.90  & 0.03  & r & NOT \\ 
122,342.8 & $2\times300$ & 21.82  & 0.12  & r & NOT \\ 
41,175.4 & $2\times120$ & 20.36  & 0.03  & i & NOT \\ 
41,873.7 & $3\times120$ & 19.92  & 0.04  & z & NOT \\ 
123,025.7 & $2\times200$ & 20.99  & 0.11  & z & NOT \\ 
205,510.8 & $9\times200$ & 21.41  & 0.13  & z & NOT \\ 
558,187.0 & $16\times180$ & 21.63 & 0.10 & z & NOT\\
36,686.  & 30. & 20.75 & 0.03 & r & GTC/OSIRIS+ \\
36,803.  & 30. & 20.70 & 0.06 & r & GTC/OSIRIS+ \\
36,943.  & 30. & 20.76 & 0.04 & r & GTC/OSIRIS+ \\
900,109. & 4$\times$250 & 22.99 & 0.04 & r & GTC/OSIRIS+ \\ 
900,873. & 15$\times$90 & 21.82 & 0.09 & z & GTC/OSIRIS+ \\ 
976,725.  & 15$\times$90 & 21.45 & 0.03 & z & GTC/OSIRIS+ \\
978,336. & 4$\times$250 & 22.51 & 0.02 & r & GTC/OSIRIS+ \\
2.61643e+06 & 30$\times$60 & 24.51 & 0.10 & u & GTC/HiPERCAM\\
2.61643e+06 & 30$\times$60 & 23.79 & 0.11 & g & GTC/HiPERCAM \\
2.61643e+06 & 30$\times$60 & 22.80 & 0.04 & r & GTC/HiPERCAM \\
2.61643e+06 & 30$\times$60 & 22.03 & 0.09 & i & GTC/HiPERCAM \\
2.61643e+06 & 30$\times$60 & 21.72 & 0.08 & z & GTC/HiPERCAM \\
2.88413e+06 & 30$\times$60 & 24.49 & 0.07 & u & GTC/HiPERCAM \\
2.88413e+06 & 30$\times$60 & 23.76 & 0.12 & g & GTC/HiPERCAM \\
2.88413e+06 & 30$\times$60 & 22.78 & 0.03 & r & GTC/HiPERCAM \\
2.88413e+06 & 30$\times$60 & 21.97 & 0.07 & i & GTC/HiPERCAM \\
2.88413e+06 & 30$\times$60 & 21.62 & 0.09 & z & GTC/HiPERCAM \\
\noalign{\smallskip}\hline \\
\end{tabular}
\end{center}
\end{table}

\begin{table}
\begin{center}
\caption[]{Asiago/REM/TNG/LBT/MISTRAL observations}\label{tab_REM}
\begin{tabular}{rccccc}
  \hline\noalign{\smallskip}
Mid time (s) &  Exposure time (s)      & Mag (AB) & Mag$_{\rm err}$ & Filter &Telescope                  \\
  \hline\noalign{\smallskip}
  20,736.0 & 3$\times$300 & 19.67 & 0.28   &  i   &  Asiago/Schmidt\\
  21,513.6 & 2$\times$30 & 20.20 & 0.08   &  r   &  TNG\\
  32,918.4 & 12$\times$240 & 20.51 & 0.30   &  r   &  REM\\
  33,177.6 & 3$\times$60 &17.87$^\ast$   &0.33     &J     &REM\\
  207,360.0  & 42$\times$20 &18.55$^\ast$   &0.08     &H   &TNG\\
  1,157,760.0 &  19$\times$30 &21.82 & 0.04    & z    &TNG\\
  6,523,200.0  &  24$\times$100 &22.68 & 0.03     &r     &LBT\\
  6,523,200.0  &  20$\times$100 &21.76 & 0.04    & z     &LBT \\
  24,660.      & $3\times300$ & 20.63 & 0.03 & r & OHP-T193/MISTRAL \\
\noalign{\smallskip}\hline \\
\end{tabular}
\tablecomments{0.86\textwidth}{All magnitudes marked with ``${\ast}$"  are in the Vega system. }
\end{center}
\end{table}

\begin{table}
\begin{center}
\caption[]{Inference result of model parameters}\label{tab_param}
\begin{tabular}{lcr}
  \hline\noalign{\smallskip}
Parameter &  Prior range      & Posterior value \\
  \hline\noalign{\smallskip}
$\log_{10}\Gamma_{0}$    & $[2,2.37]$  & $2.37^{+0.02}_{-0.02}$ \\
$\log_{10}\epsilon_{e,f}$  & $[-5,-0.1]$   & $-1.00^{+0.12}_{-0.14}$ \\
$\log_{10}\epsilon_{B,f}$  & $[-8,-0.1]$   & $-6.41^{+0.39}_{-0.31}$ \\
$\log_{10}\theta_{j}$  (rad)    & $[-1,0.4]$   & $-0.66^{+0.11}_{-0.11}$ \\
$\log_{10}E_{\rm k,iso}$ (erg) & $[53,56.23]$  & $54.72^{+0.16}_{-0.16}$ \\
$p_{f}$             & $[2,3]$   &  $2.41^{+0.03}_{-0.03}$ \\
\hline
$\log_{10}\epsilon_{e,r}$  & $[-5,-0.1]$ & $-1.27^{+0.28}_{-0.34}$ \\
$\log_{10}\epsilon_{B,r}$  & $[-8,-0.1]$ & $-3.93^{+0.45}_{-0.48}$ \\
$p_{r}$           & $[2,3]$ & $2.08^{+0.08}_{-0.04}$ \\
\hline
$\log_{10}n_{0}$      & $[-3,3]$ & $0.05^{+0.18}_{-0.18}$ \\
$E(B-V)$$_{\rm host}$  & $[0.3,0.6]$ & $0.37^{+0.02}_{-0.02}$\\
$\log_{10}f_{sys}$      & $[-2,0]$ & $-0.51^{+0.02}_{-0.02}$ \\
\noalign{\smallskip}\hline \\
\end{tabular}
\end{center}
\tablecomments{0.86\textwidth}{Uniform prior distribution.}
\end{table}

\FloatBarrier

\label{lastpage}
\newpage
\clearpage
\bibliography{ms2025-0222}{}
\bibliographystyle{raa}

\end{document}